\documentclass[
reprint,
superscriptaddress,
amsmath,
amssymb,
aps,
prb,
floatfix,
]{revtex4-2}

\usepackage{graphicx}
\usepackage{dcolumn}
\usepackage{bm}
\usepackage[T1]{fontenc}
\usepackage[utf8]{inputenc}
\usepackage{lmodern}
\usepackage[percent]{overpic}
\usepackage[
colorlinks=true,
linkcolor=blue,
citecolor=blue,
urlcolor=blue
]{hyperref}
\usepackage[mathlines]{lineno}
\usepackage{siunitx}
\usepackage[caption=false]{subfig}
\usepackage{upgreek}
\usepackage{balance}

\graphicspath{{.}{figs/}}

\begin{document}

	\title{Modification of optical properties in ultrathin ALD-grown NbN layers using oxygen plasma treatment}
	
	
	\author{Mustafa Teke}
	\email{mustafa.teke@leibniz-ipht.de}
	\affiliation{
		Leibniz Institute of Photonic Technology,
		07745 Jena, Germany
	}
	\affiliation{
		Institute for Solid State Physics,
		Friedrich Schiller University Jena,
		07743 Jena, Germany
	}
	
	\author{Nitesh Yerra}
	\affiliation{
		Leibniz Institute of Photonic Technology,
		07745 Jena, Germany
	}
	\affiliation{
		Institute for Solid State Physics,
		Friedrich Schiller University Jena,
		07743 Jena, Germany
	}
	
	\author{Alireza Sadeghi}
	\affiliation{
		Leibniz Institute of Photonic Technology,
		07745 Jena, Germany
	}
	\affiliation{
		Institute for Solid State Physics,
		Friedrich Schiller University Jena,
		07743 Jena, Germany
	}
	
	\author{Mario Ziegler}
	\affiliation{
		Leibniz Institute of Photonic Technology,
		07745 Jena, Germany
	}
	
	\author{Detlef Born}
	\affiliation{
		Leibniz Institute of Photonic Technology,
		07745 Jena, Germany
	}
	
	\author{Nan Du}
	\affiliation{
		Leibniz Institute of Photonic Technology,
		07745 Jena, Germany
	}
	
	\author{Marco Diegel}
	\affiliation{
		Leibniz Institute of Photonic Technology,
		07745 Jena, Germany
	}
	
	\author{Uwe H\"ubner}
	\affiliation{
		Leibniz Institute of Photonic Technology,
		07745 Jena, Germany
	}

	\author{Sahitya V. Vegesna}
	\email{sahityav.vegesna@leibniz-ipht.de}
	\affiliation{
		Leibniz Institute of Photonic Technology,
		07745 Jena, Germany
	}
	\affiliation{
		Institute for Solid State Physics,
		Friedrich Schiller University Jena,
		07743 Jena, Germany
	}
	
	\author{Heidemarie Kr\"uger}
	\email{heidemarie.krueger@leibniz-ipht.de}
	\affiliation{
		Leibniz Institute of Photonic Technology,
		07745 Jena, Germany
	}
	\affiliation{
		Institute for Solid State Physics,
		Friedrich Schiller University Jena,
		07743 Jena, Germany
	}
	
	\date{\today}
	
	\begin{abstract}
		NbN thin films with a nominal thickness of 10\,nm were grown by atomic layer deposition (ALD) on Al$_2$O$_3$-buffered Si substrates and treated with oxygen plasma for 0\,s (untreated), 20\,s, 40\,s, and 60\,s to investigate oxygen-induced changes in their optical properties. Spectroscopic ellipsometry (SE) measurements were performed before and after oxygen plasma treatment over the ultraviolet--visible--near-infrared (1.24--6.20\,eV) and infrared (0.041--0.827\,eV) spectral ranges. Film thickness determined by X-ray reflectivity (XRR) was used as a fixed parameter in the optical modeling to extract the dielectric function, optical constants, and effective free-carrier concentration. We observed the emergence of a distinct absorption feature near 4.9\,eV following oxygen plasma treatment, consistent with the oxidation of the NbN surface and the resulting change in its optical properties. Increasing the treatment time from 0 to 60\,s reduced the effective carrier concentration from $1.20 \times 10^{22}$ to $1.03 \times 10^{22}$\,cm$^{-3}$ and the effective carrier mobility from 2.44 to 2.27\,cm$^2$\,V$^{-1}$\,s$^{-1}$. The obtained results provide a basis for understanding the effects of oxygen plasma treatment on NbN thin films.

	\end{abstract}
	\maketitle
	\section{Introduction}
	
	Niobium nitride (NbN) is a transition-metal nitride distinguished by its electrical, optical, and superconducting properties \cite{Kobayashi2025}, which make it attractive for superconducting nanowire single-photon detectors (SNSPDs) \cite{Natarajan2012,Hadfield2005}, hot-electron bolometer mixers \cite{Ren2019,Shurakov2015}, and superconducting microwave resonators \cite{Frasca2023,Abdo2006}. High-quality NbN films exhibit superconducting transition temperatures ($T_{\mathrm{c}}$) around 16\,K \cite{Knau1991,Shapoval2011,Hazra2016}, critical current densities ($J_{\mathrm{c}}$) exceeding $10^{7}$\,A/cm$^{2}$ \cite{Boffa1996,Jia2014,Cheng2020}, low microwave surface resistance \cite{Oates1991}, and strong optical absorption from the visible to the infrared spectral range \cite{Isakov2018}. Since these properties are highly sensitive to film thickness, microstructure, and composition \cite{Hazra2016,Licata2022,Kalal2021}, precise control over the deposition process is particularly important for ultrathin NbN films. Atomic layer deposition (ALD) offers atomic-scale thickness control through self-limiting surface reactions, together with high thickness uniformity and conformal coverage \cite{Parsons2011, Getnet2025}. These characteristics make ALD particularly suitable for the controlled growth of ultrathin NbN films, where small variations in thickness or composition can substantially affect their superconducting and optical properties \cite{Linzen2017,Semenov2009}.

	In ultrathin NbN films, surface oxidation can reduce the effective thickness of the superconducting NbN layer and alter the composition of the near-surface region \cite{Licata2022,Ermolieff1985,Halbritter1987}. Ermolieff \textit{et al.} reported the formation of Nb$_2$O$_5$ as the predominant surface oxide, with NbO and NbO$_2$ located at the interface between Nb$_2$O$_5$ and NbN \cite{Ermolieff1985}. Further X-ray photoelectron spectroscopy (XPS) measurements showed that oxidation initially involves oxygen diffusion into NbN, followed by increasingly pronounced Nb--O bonding with increasing oxygen exposure \cite{Ermolieff1986}. Depth-resolved XPS studies have also revealed changes in the phase composition beneath the surface oxide \cite{Lubenchenko2017}. In oxygen-plasma-treated ALD NbN films, XPS measurements identified a Nb$_2$O$_5$-rich surface layer, with no detectable change in the chemical composition of the probed film interior relative to the untreated film \cite{Yerra2026}. Such surface modifications may alter the effective free-carrier response and complex dielectric function of the film, thereby influencing the coupling and absorption of incident light. These observations motivate investigating how controlled surface oxidation modifies the optical properties of ultrathin NbN films, with particular relevance to photon absorption in NbN-based SNSPDs operated below the superconducting transition temperature $T_c$ \cite{Hadfield2009}. Understanding how oxidation influences the optical and electronic properties of NbN requires a characterization technique sensitive to both free-carrier response and interband transitions \cite{Aspnes2014,Politano2023}. Spectroscopic ellipsometry (SE) is well suited for this purpose, providing non-destructive access to the complex dielectric function over a broad spectral range \cite{Politano2023}. In this work, we investigate oxygen-plasma-induced modifications in the optical and electronic response of ALD-grown NbN thin films using SE.
	
	\section{Experimental Methods}
	
	\subsection{Film deposition and plasma treatment}
	
\begin{figure}[t]
		\centering
		\includegraphics[width=1\columnwidth]{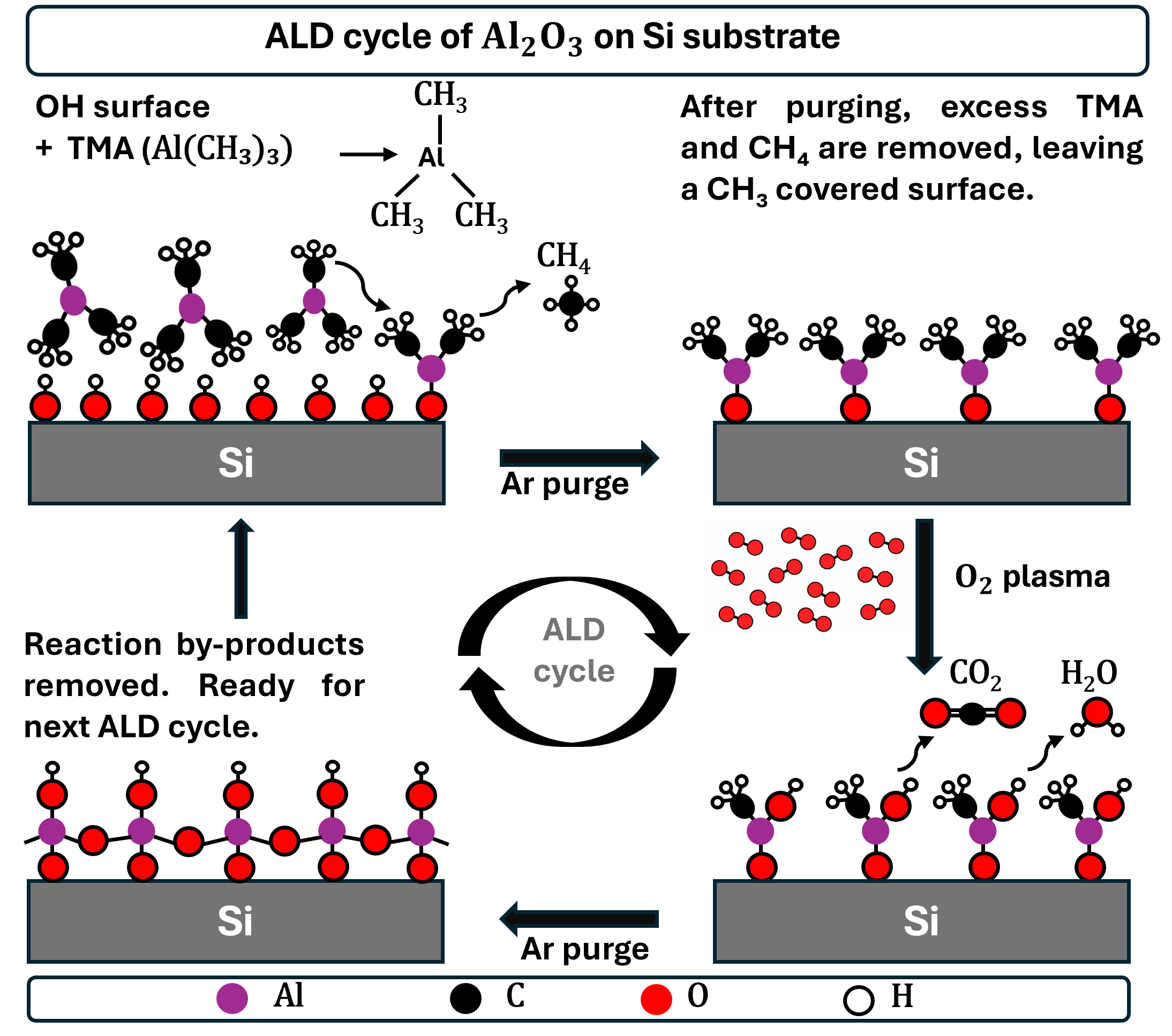}
		\caption{One ALD cycle for Al$_2$O$_3$ growth. Following the TMA pulse, surface hydroxyl groups react with TMA to form surface Al--CH$_3$ species while releasing CH$_4$. The subsequent Ar purge removes excess TMA and reaction byproducts before O$_2$ plasma converts the surface methyl groups into Al--O bonds, releasing CO$_2$ and H$_2$O. A second Ar purge removes the remaining reaction byproducts, thereby completing the deposition cycle.}
		
		\label{fig:AL2O3_ALD}
	\end{figure}

	A total of four $10\times10$\,mm$^{2}$ substrates were prepared from a commercially available 4-inch Czochralski-grown, boron-doped p-type Si(100) wafer with a resistivity of 1--30\,$\Omega$\,cm. A 4-nm-thick Al$_2$O$_3$ buffer layer, which served as an etch-resistant layer for the following NbN deposition, was subsequently deposited on each substrate by plasma-enhanced atomic layer deposition (PEALD) using an Oxford Plasma Technology OpAL reactor, forming the Al$_2$O$_3$/Si substrate stack employed in this work. Fig.~\ref{fig:AL2O3_ALD} illustrates the sequence of surface reactions occurring in one Al$_2$O$_3$ ALD cycle. Each cycle consisted of a 50\,ms trimethylaluminum (TMA, Al(CH$_3$)$_3$) pulse, a 4500\,ms Ar purge, a 3000\,ms O$_2$ plasma exposure, and a final 2000\,ms Ar purge. The depositions were carried out at a substrate temperature of 150\,$^\circ$C using an O$_2$ plasma power of 300\,W, resulting in a growth per cycle (GPC) of 1.32\,\AA/cycle. Following the TMA pulse, TMA reacts with the hydroxyl groups on the substrate surface, leaving surface Al--CH$_3$ and Al--(CH$_3$)$_2$ species while CH$_4$ is released. An Ar purge removes excess TMA together with the reaction byproducts. The O$_2$ plasma subsequently oxidizes the surface CH$_3$ groups, forming Al--O bonds while releasing CO$_2$ and H$_2$O. A second Ar purge removes the remaining CO$_2$ and H$_2$O, along with residual volatile reaction byproducts, before the next TMA pulse, thereby completing one ALD cycle.

	\begin{figure*}
		\centering
		\includegraphics[width=0.9\textwidth]{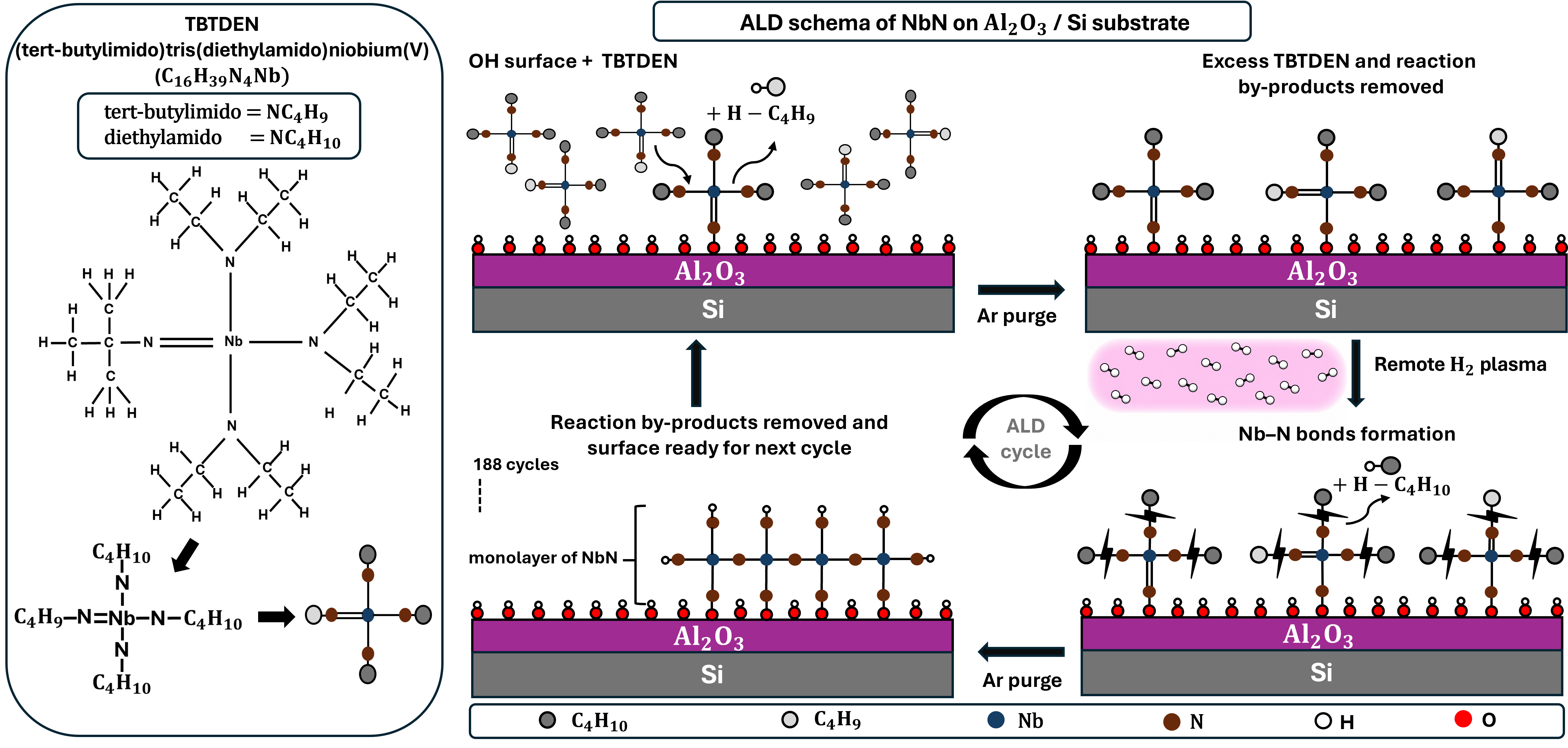}
		\caption{Schematic illustration of the NbN deposition process by PEALD using the TBTDEN precursor and remote H$_2$ plasma. Following the TBTDEN pulse, precursor molecules adsorb on the substrate surface. An Ar purge removes excess precursor and reaction products before remote H$_2$ plasma removes the organic ligands and promotes Nb--N bond formation. A final Ar purge removes residual reaction products and prepares the surface for the subsequent PEALD cycle. The depicted reaction mechanism is based on our assumption and has not been experimentally verified. For clarity, the atomic configurations and bonds are shown in 2D, although the resulting NbN film has a 3D structure with more complex atomic coordination.}
		\label{fig:NbN_ALD}
	\end{figure*} 
	
	\begin{figure}[t]
		\centering
		\includegraphics[width=1\columnwidth]{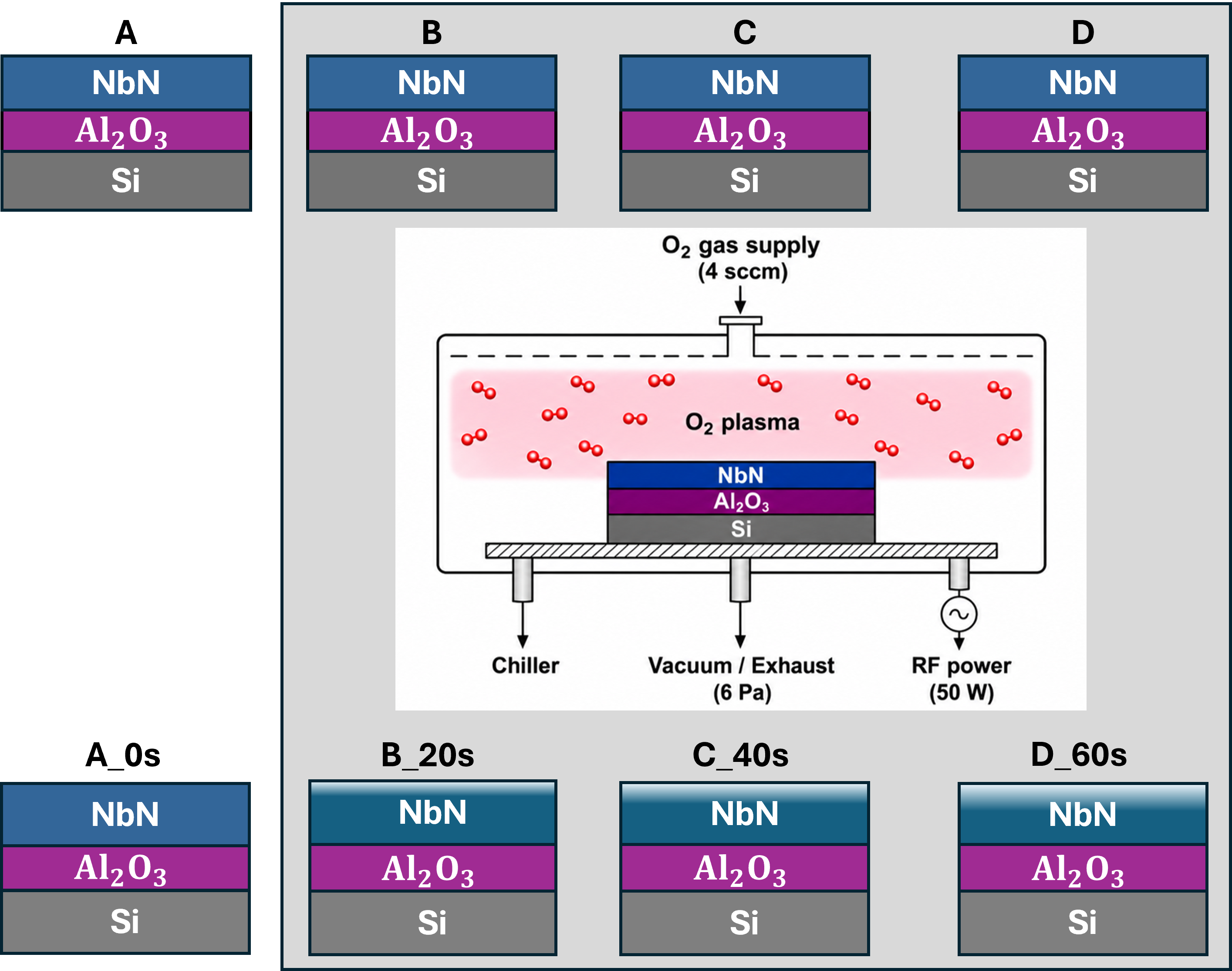}
		\caption{Experimental workflow of the oxygen plasma treatment. SE measurements were performed before and after oxygen plasma treatment. Before treatment, the NbN/Al$_2$O$_3$/Si samples were labeled A, B, C, and D. After treatment, the samples were renamed A\_0s, B\_20s, C\_40s, and D\_60s according to oxygen plasma treatment times of 0, 20, 40, and 60\,s, respectively.}
		\label{fig:plasma_treatment}
	\end{figure}
	
	Subsequently, NbN films with a nominal thickness of 10\,nm were deposited on the Al$_2$O$_3$/Si substrate stack using PEALD (Fig.~\ref{fig:NbN_ALD}). All four samples were grown during the same PEALD run to ensure identical growth conditions. Deposition was carried out in the same Oxford Plasma Technology OpAL reactor equipped with an inductively coupled remote-plasma source. In contrast to the standard reactor configuration, the system incorporated a two-stage dry pumping system, providing a base pressure of approximately 2\,mTorr. The reactor was operated inside a nitrogen-filled glovebox to minimize contamination from atmospheric moisture and oxygen. During deposition, the substrate temperature was maintained at $380\,^{\circ}$C and the chamber pressure was controlled between 5 and 16\,mTorr. (tert-Butylimido)-tris(diethylamino)niobium (TBTDEN, Heraeus Precious Metals GmbH, Germany) was used as the Nb precursor, while H$_2$ plasma served as the co-reactant. Fig.~\ref{fig:NbN_ALD} shows the individual steps of one NbN PEALD cycle. Each cycle consisted of a 4\,s TBTDEN exposure, followed by a 7\,s Ar purge, a 60\,s H$_2$ plasma exposure, and a second 10\,s Ar purge. During the precursor pulse, TBTDEN molecules chemisorb on the available reactive surface sites until saturation is reached, forming a self-limiting Nb-containing precursor layer. The subsequent Ar purge eliminates excess precursor molecules together with volatile reaction byproducts, suppressing gas-phase reactions that could otherwise lead to chemical vapor deposition. Exposure to H$_2$ plasma removes the organic ligands from the chemisorbed precursor while converting the Nb-containing surface layer into NbN. The second Ar purge clears the remaining gaseous reaction byproducts before the next deposition cycle. Completion of one PEALD cycle results in the deposition of a single NbN monolayer with an average GPC of approximately 0.53\,\AA/cycle. Consequently, 188 PEALD cycles were required to obtain the nominal NbN thickness of 10\,nm. Following NbN deposition, the four samples were labeled A, B, C, and D and subjected to oxygen plasma treatment for 0 (no treatment), 20, 40, and 60\,s, respectively. The oxygen plasma treatment was carried out in a SENTECH Instruments SI~500-408 reactive-ion etching chamber operated in a capacitively coupled plasma configuration, as shown in Fig.~\ref{fig:plasma_treatment}. Oxygen was introduced into the chamber at a flow rate of 4\,sccm, while the chamber pressure and radio-frequency (RF) power were maintained at 6\,Pa and 50\,W. After oxygen plasma treatment, the samples were designated as A\_0s, B\_20s, C\_40s, and D\_60s according to their respective treatment times.
	
	\newpage
	
	\subsection{Optical and structural characterization}

	Spectroscopic ellipsometry was used to determine the dielectric response of NbN thin films before and after oxygen plasma treatment. UV--Vis measurements (0.2--1.0\,\si{\micro\meter}) were performed using a rotating compensator ellipsometer (RC2, J.A.~Woollam Co.) \cite{JAWoollam2025}, providing high sensitivity to thickness variations in thin films due to strong interference and absorption effects at shorter wavelengths \cite{Hilfiker2008}. Infrared measurements (1.5--30\,\si{\micro\meter}) were carried out using an IR-VASE Mark II ellipsometer \cite{JAWoollam2025}, primarily probing the free-carrier and low-energy dielectric response \cite{Kasic2002}. The measurements were performed at incidence angles from $45^\circ$ to $70^\circ$ in $5^\circ$ steps.
	
	The NbN and Al$_2$O$_3$ layer thicknesses were determined after oxygen plasma treatment by X-ray reflectivity (XRR) using a Malvern Panalytical MPD Pro theta--theta diffractometer. The reflectivity curves were recorded in the $\Upomega$--$2\uptheta$ geometry over an angular range of $0.04^\circ$--$3.5^\circ$, with a step size of $0.005^\circ$ and an acquisition time of 5\,s per step. Before each measurement, the sample height was adjusted until the direct-beam intensity was reduced to 50\% of its initial value.
	
	\section{Results}
	
	\subsection{X-ray reflectivity analysis}
	
	\begin{figure*}[t]
		\centering
		\nolinenumbers
		
		\resizebox{0.95\textwidth}{!}{%
			\begin{minipage}{\textwidth}
				\centering
				
				\begin{minipage}{0.490\textwidth}
					\centering
					\includegraphics[width=\linewidth]{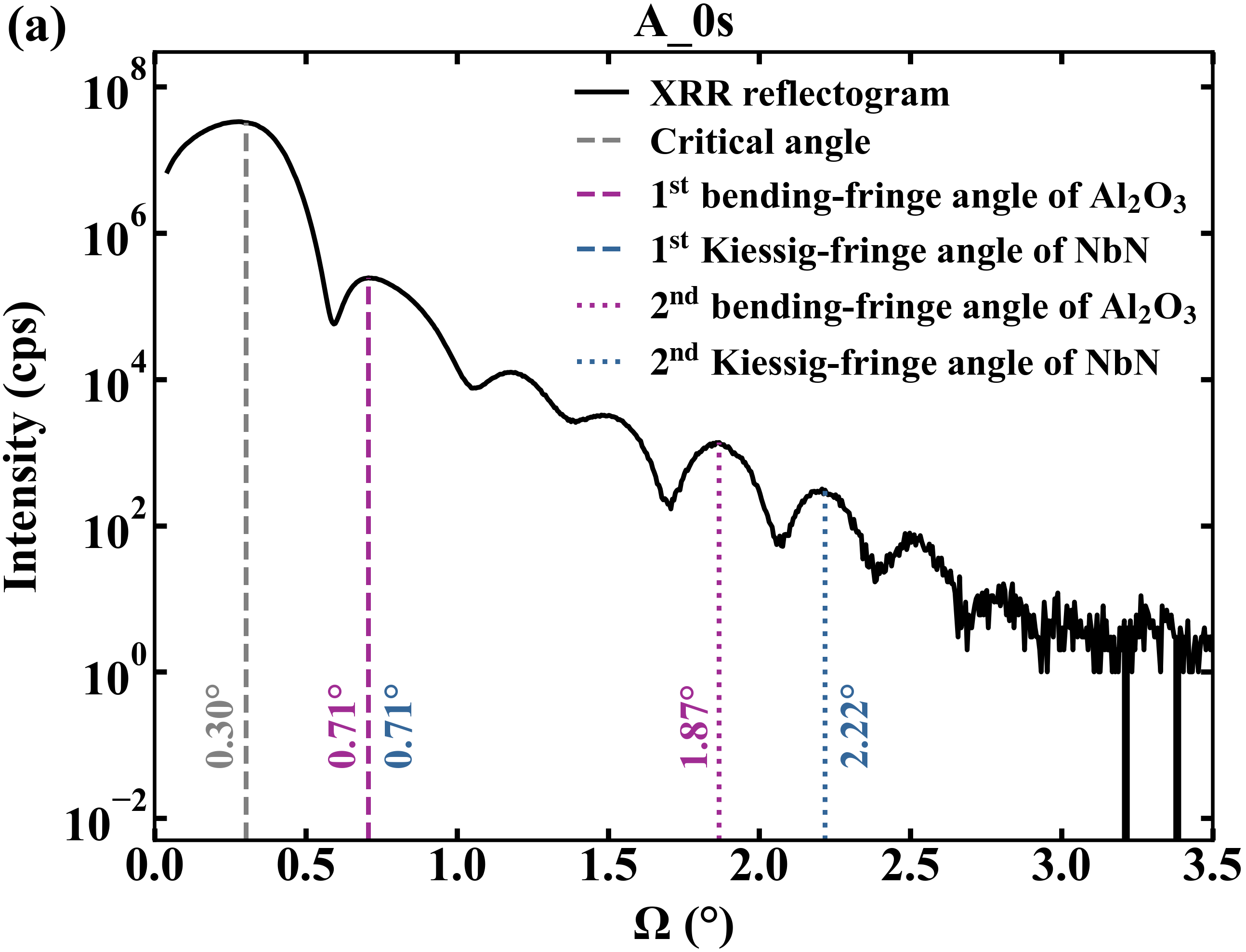}
				\end{minipage}\hfill
				\begin{minipage}{0.490\textwidth}
					\centering
					\includegraphics[width=\linewidth]{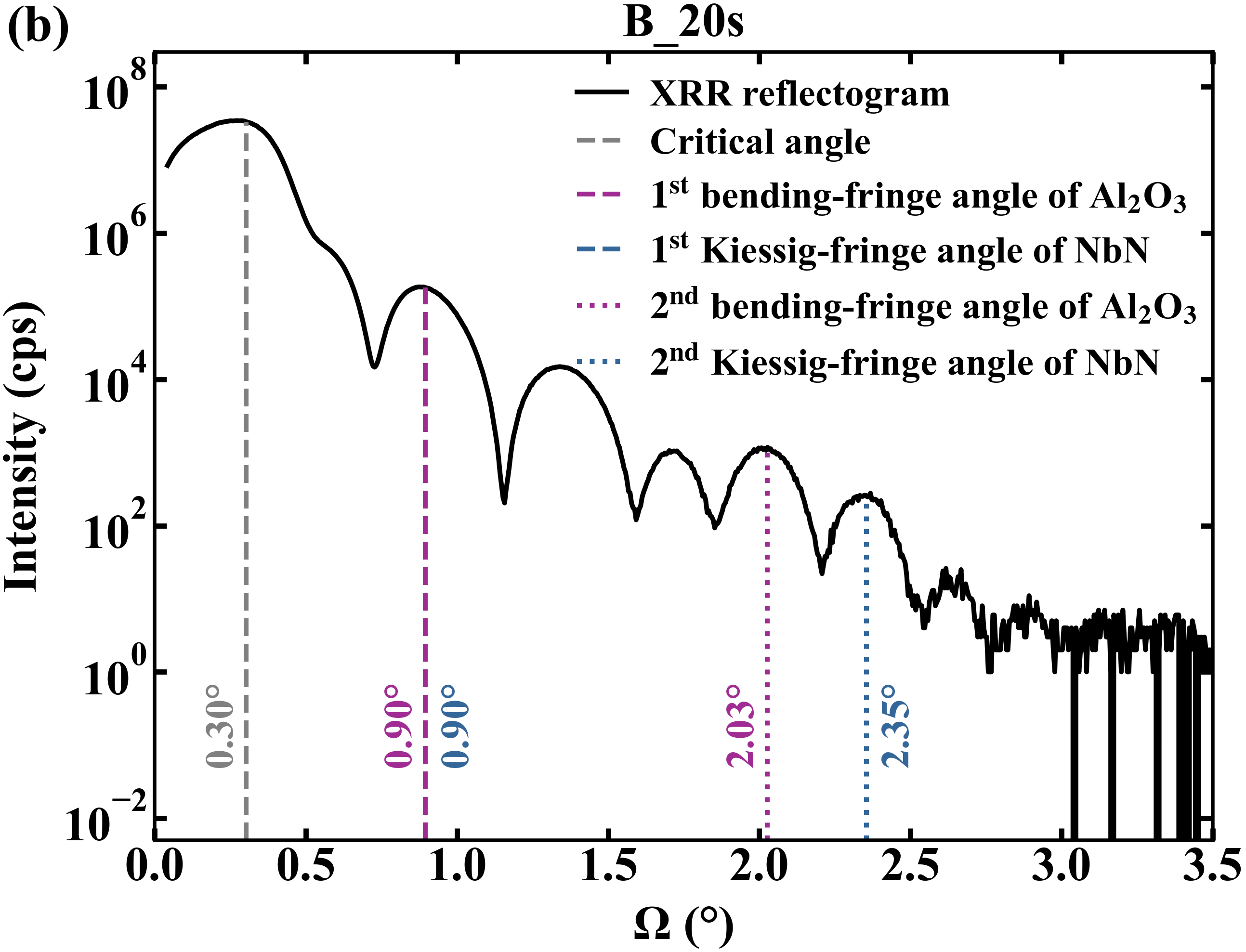}
				\end{minipage}
				
				\vspace{0.25cm}
				
				\begin{minipage}{0.490\textwidth}
					\centering
					\includegraphics[width=\linewidth]{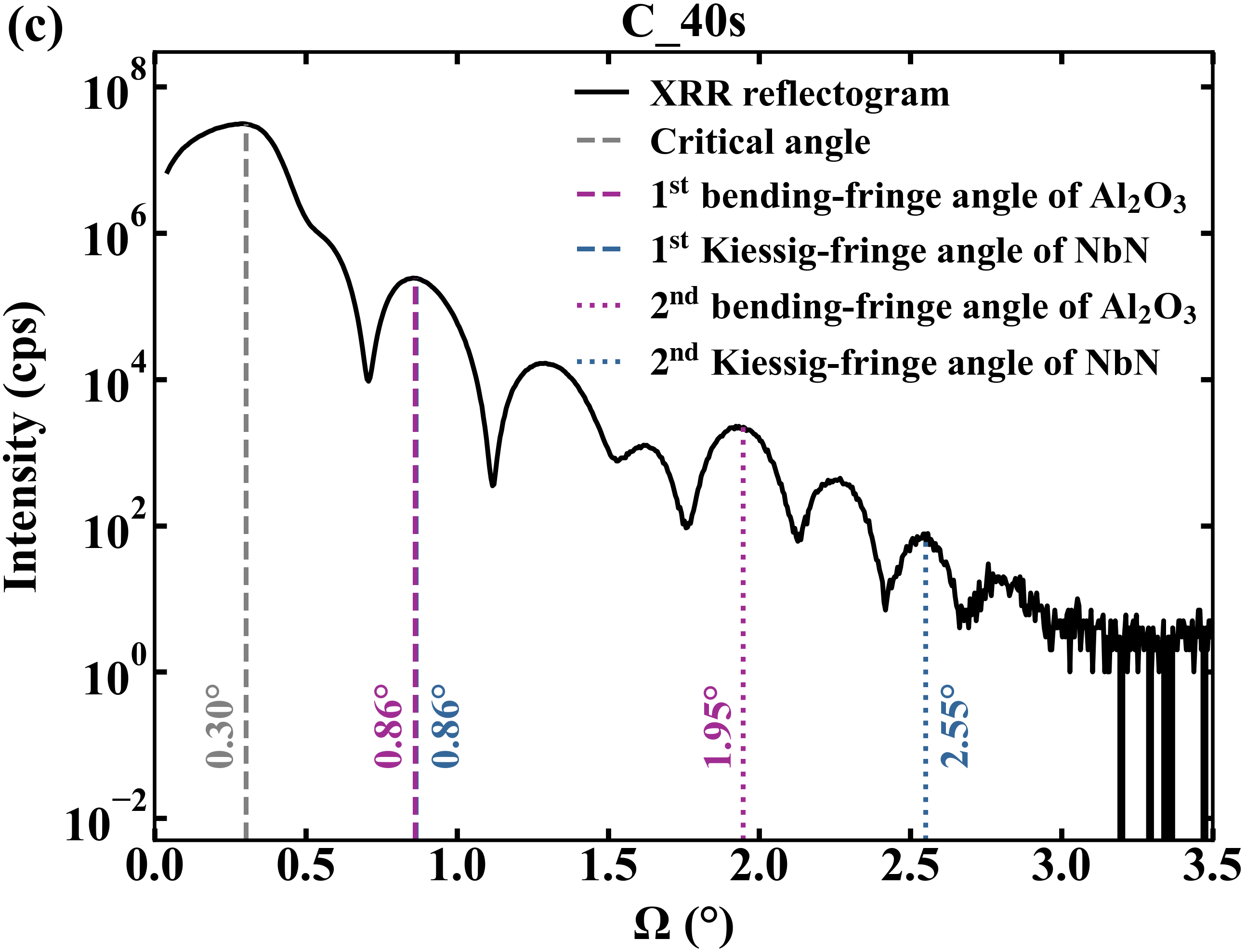}
				\end{minipage}\hfill
				\begin{minipage}{0.490\textwidth}
					\centering
					\includegraphics[width=\linewidth]{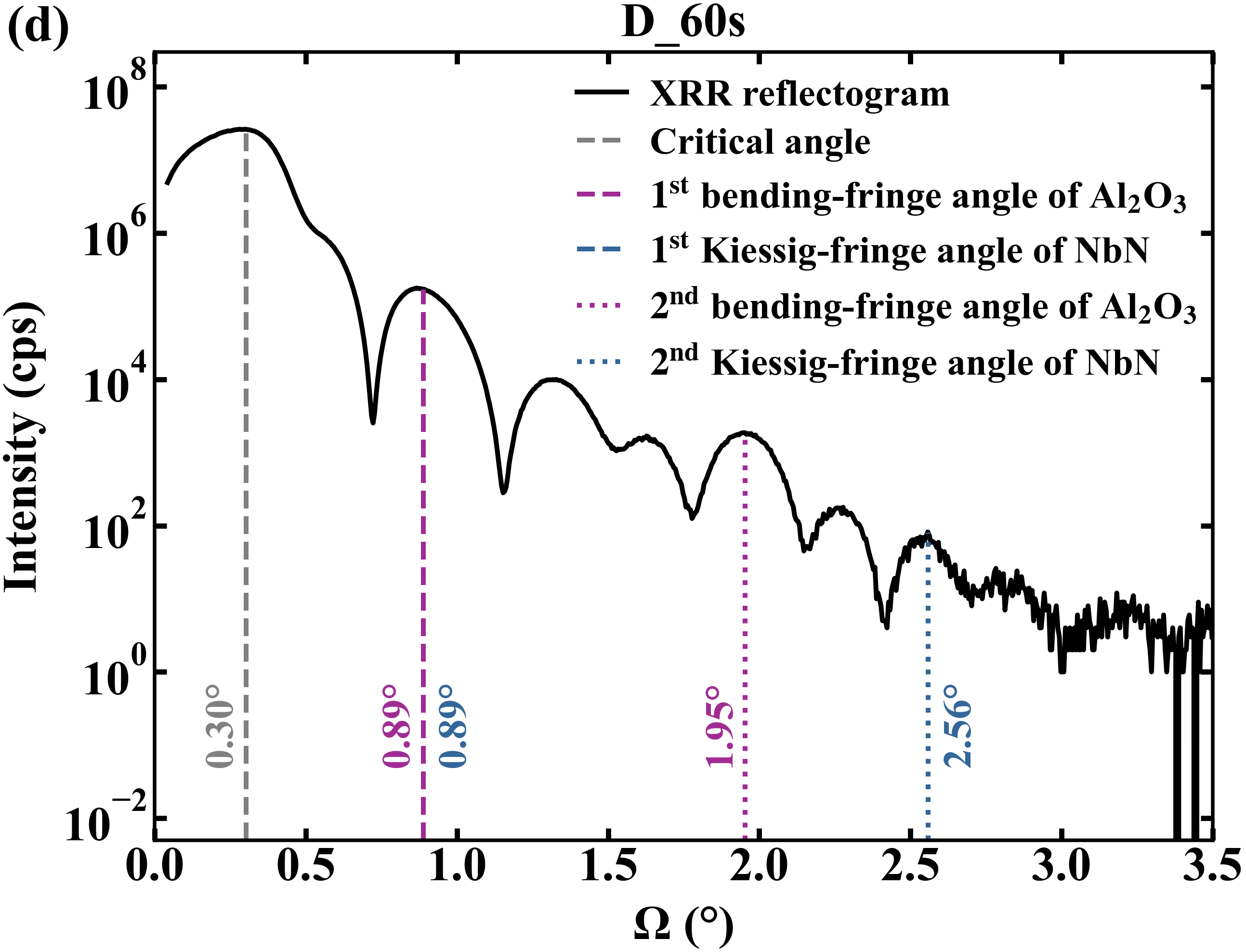}
				\end{minipage}
				
			\end{minipage}%
		}
		
		\caption{XRR reflectograms of NbN thin films deposited on Al$_2$O$_3$/Si substrates after oxygen plasma treatment for (a) 0\,s, (b) 20\,s, (c) 40\,s, and (d) 60\,s. The gray dashed line indicates the critical angle, corresponding to the onset of total external reflection. The blue dashed and dotted lines mark the $1^{\mathrm{st}}$ and $2^{\mathrm{nd}}$ interference-fringe angles of the NbN layer, while the purple dashed and dotted lines denote the $1^{\mathrm{st}}$ and $2^{\mathrm{nd}}$ bending-fringe angles of the Al$_2$O$_3$ layer. The thicknesses of the NbN and Al$_2$O$_3$ layers were determined from the angular spacing between the corresponding characteristic angles using the XRR interference relations, with the critical angle serving as the reference for the analysis.}
		\label{fig:xrr}
	\end{figure*}
	
	 The measured XRR reflectograms were analyzed using Malvern Panalytical AMASS (version 1.1), with the layer thicknesses determined by the fringe-thickness method. For each reflectogram, the critical angle and characteristic fringe positions were identified visually and selected manually, as indicated in Fig.~\ref{fig:xrr}. The critical angle, $\uptheta_c$, marks the onset of significant X-ray penetration into the sample. Above $\uptheta_c$, X-rays reflected from the top surface of the NbN film interfere with X-rays reflected from the underlying interfaces, producing the periodic maxima and minima observed in the XRR curve as Kiessig fringes. Their angular spacing is inversely related to the layer thickness, providing the basis for the thickness determination. For the NbN layer, two well-defined Kiessig-fringe maxima separated by four or five fringe periods, depending on the sample, were selected. The thickness corresponding to a single fringe period was calculated from their angular positions and the critical angle, and the total NbN thickness was obtained by multiplying this value by the number of periods between the selected positions. In the NbN/Al$_2$O$_3$ bilayer structure, the thinner Al$_2$O$_3$ layer produces a slower modulation of the Kiessig-fringe pattern, observed as a bending of the fringes. Two successive maxima of this modulation were selected, with their angular separation corresponding to one complete period used to determine the Al$_2$O$_3$ thickness. The calculated layer thicknesses are sensitive to the angular positions of the selected fringes. Small variations in these positions, associated with their visual identification and manual selection, can introduce differences of a few~\AA{} in the determined thicknesses. For sample D\_60s, the critical angle was identified at $0.3026^\circ$, while the selected Kiessig-fringe maxima for the NbN layer were located at $0.8881^\circ$ and $2.5587^\circ$ (see Fig.~\hyperref[fig:xrr]{\ref*{fig:xrr}(d)}). These positions span five fringe periods, corresponding to a thickness of 12.9\,nm for the NbN layer. For the Al$_2$O$_3$ layer, the selected bending-fringe maxima at $0.8881^\circ$ and $1.9538^\circ$ define one complete modulation period, from which a thickness of 4.0\,nm was determined. In addition to the NbN and Al$_2$O$_3$ layers, a thin native SiO$_2$ layer was identified between the Si substrate and the Al$_2$O$_3$ layer in all four samples. Such oxide layers naturally form on Si surfaces through reactions with oxygen and moisture upon exposure to ambient air \cite{Morita1990}. The SiO$_2$ thicknesses obtained from the XRR analysis ranged from 1.8 to 1.9\,nm, consistent with the nanometer-scale native oxide thicknesses reported in the literature \cite{Hafshejani2023,Li2026}. The layer thicknesses obtained for all samples are summarized in Table~\ref{tab:xrr_thickness}. Since sample A\_0s was not treated with oxygen plasma, its measured NbN, Al$_2$O$_3$, and native SiO$_2$ thicknesses were taken as the reference values before treatment. As all four samples were prepared in the same deposition run under identical growth conditions, the corresponding layer thicknesses of samples B\_20s, C\_40s, and D\_60s before treatment were assumed to be equal to those of sample A\_0s.
	\begin{table}[t]
		\centering
		\caption{NbN, Al$_2$O$_3$, and native SiO$_2$ layer thicknesses determined by XRR for samples treated with oxygen plasma for 0, 20, 40, and 60\,s. The error values reflect possible variations in the calculated layer thicknesses due to the visual identification and manual selection of the fringe positions.}
		\label{tab:xrr_thickness}
		
		\renewcommand{\arraystretch}{1.15}
		
		\begin{tabular*}{\columnwidth}{@{\extracolsep{\fill}}lccc@{}}
			\hline\hline
			Sample & NbN (nm) & Al$_2$O$_3$ (nm) & SiO$_2$ (nm) \\
			\hline
			A\_0s  & $11.4 \pm 0.3$ & $3.7 \pm 0.3$ & $1.8 \pm 0.1$ \\
			B\_20s & $11.8 \pm 0.3$ & $3.8 \pm 0.3$ & $1.9 \pm 0.1$ \\
			C\_40s & $12.8 \pm 0.3$ & $3.9 \pm 0.3$ & $1.9 \pm 0.1$ \\
			D\_60s & $12.9 \pm 0.3$ & $4.0 \pm 0.3$ & $1.9 \pm 0.1$ \\
			\hline\hline
		\end{tabular*}
	\end{table}

	\subsection{Spectroscopic ellipsometry modeling}
	
	The measured ellipsometric parameters $\Psi$ and $\Delta$ for samples A\_0s, B\_20s, C\_40s, and D\_60s after oxygen plasma treatment are shown in Fig.~\ref{fig:psi_delta_after}. A detailed analysis of the corresponding measurements before treatment is provided in Appendix~\ref{app:before}. The parameters $\Psi$ and $\Delta$ describe the amplitude and phase of the function $\rho$, where $\rho$ is the ratio of the reflected $p$- and $s$-polarized light intensities (Eq.~\ref{eq:rho}) \cite{Wee2022}. The complex reflectance ratio $\rho$ is given by
	\begin{equation}
		\begin{aligned}
			\rho &= \frac{r_p}{r_s}
			= \tan(\Psi)e^{i\Delta} \\
			&=
			\frac{r_{01,p}+r_{12,p}e^{-i2\beta}}
			{1+r_{01,p}r_{12,p}e^{-i2\beta}}
			\Bigg/
			\frac{r_{01,s}+r_{12,s}e^{-i2\beta}}
			{1+r_{01,s}r_{12,s}e^{-i2\beta}} .
		\end{aligned}
		\label{eq:rho}
	\end{equation}
	where $r_p$ and $r_s$ denote the total complex reflection coefficients for $p$- and $s$-polarized light, respectively \cite{Hilfiker2022}. The quantities $r_{01}$ and $r_{12}$ are the Fresnel reflection coefficients at the air$_{0}$/film$_{1}$ and film$_{1}$/substrate$_{2}$ interfaces, while $\beta$ is the phase accumulated during one-way propagation through the film \cite{Wee2022}. The measured $\Psi$ and $\Delta$ were modeled in CompleteEASE (J.A. Woollam Co.) with the multilayer structure consisting of the Si substrate, native SiO$_2$, Al$_2$O$_3$, and NbN layers. Literature-based optical data were used for the Si substrate, native SiO$_2$, and Al$_2$O$_3$ layers \cite{Herzinger1998}. The NbN, Al$_2$O$_3$, and native SiO$_2$ layer thicknesses determined by XRR were fixed in the optical model. Following oxygen plasma treatment, the oxygen-modified surface region and the underlying NbN were modeled as a single effective NbN layer with isotropic optical properties, rather than as separate optical layers. The optical response of this layer was described using a Drude--Lorentz dispersion model.
	
	The Kramers--Kronig-consistent Drude--Lorentz optical model agrees well with the measured $\Psi$ and $\Delta$ spectra acquired at incidence angles of $45^\circ$, $50^\circ$, $55^\circ$, $60^\circ$, $65^\circ$, and $70^\circ$, as shown in Fig.~\ref{fig:psi_delta_after}. The spectral region between 0.82 and 1.24\,eV, shaded in gray in Fig.~\ref{fig:psi_delta_after}, was not measured directly because it lies in the gap between the two measured spectral ranges. The curves within this region were obtained by interpolation within the optical model while maintaining Kramers--Kronig consistency, which requires the real and imaginary parts of the dielectric function to remain physically related rather than vary independently \cite{Klimchitskaya2007,Budai2014}. Minor deviations between the modeled and measured spectra become more apparent with increasing oxygen plasma treatment time, particularly at the highest incidence angles of $65^\circ$ and $70^\circ$. At large incidence angles, the ellipsometric parameters $\Psi$ and $\Delta$ become increasingly sensitive to small variations in film thickness, surface roughness, and interface profile \cite{Zeidler1974,Johs2004}. Oxygen plasma treatment is expected to alter primarily the near-surface region of the NbN films \cite{Lubenchenko2017,Ermolieff1985}, which may result in a gradual variation in optical properties normal to the surface. Since the present model represents the oxygen-modified region and the underlying NbN as a single layer, these depth-dependent variations cannot be fully captured, leading to the small residual offsets observed at the highest incidence angles.
	
	\begin{figure*}[t] 
		\centering 
		\nolinenumbers
		
		\resizebox{0.95\textwidth}{!}{%
			\begin{minipage}{\textwidth}
				\centering
				
				\begin{minipage}{0.490\textwidth} 
					\centering 
					\includegraphics[width=\linewidth]{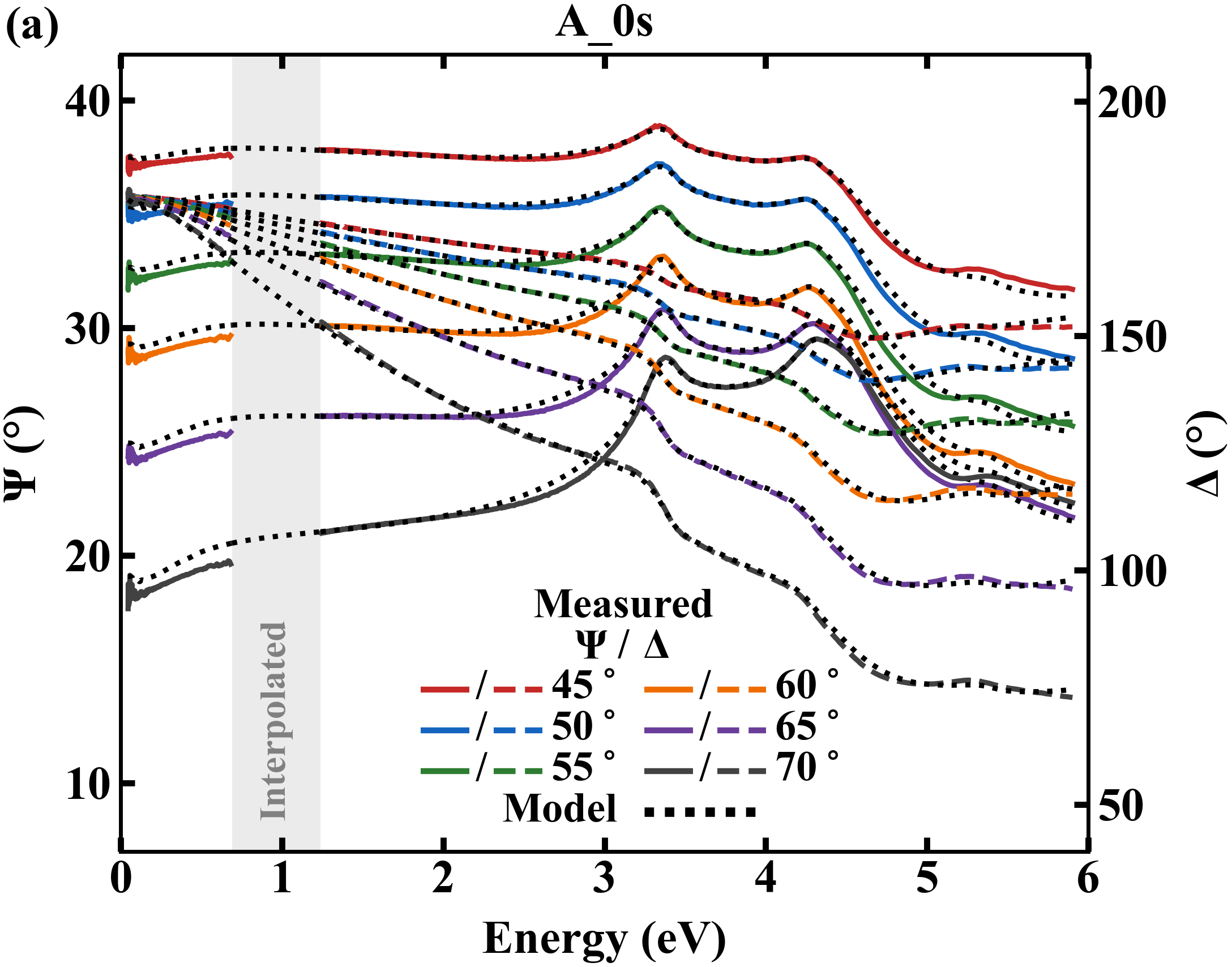} 
				\end{minipage}\hfill 
				\begin{minipage}{0.490\textwidth} 
					\centering 
					\includegraphics[width=\linewidth]{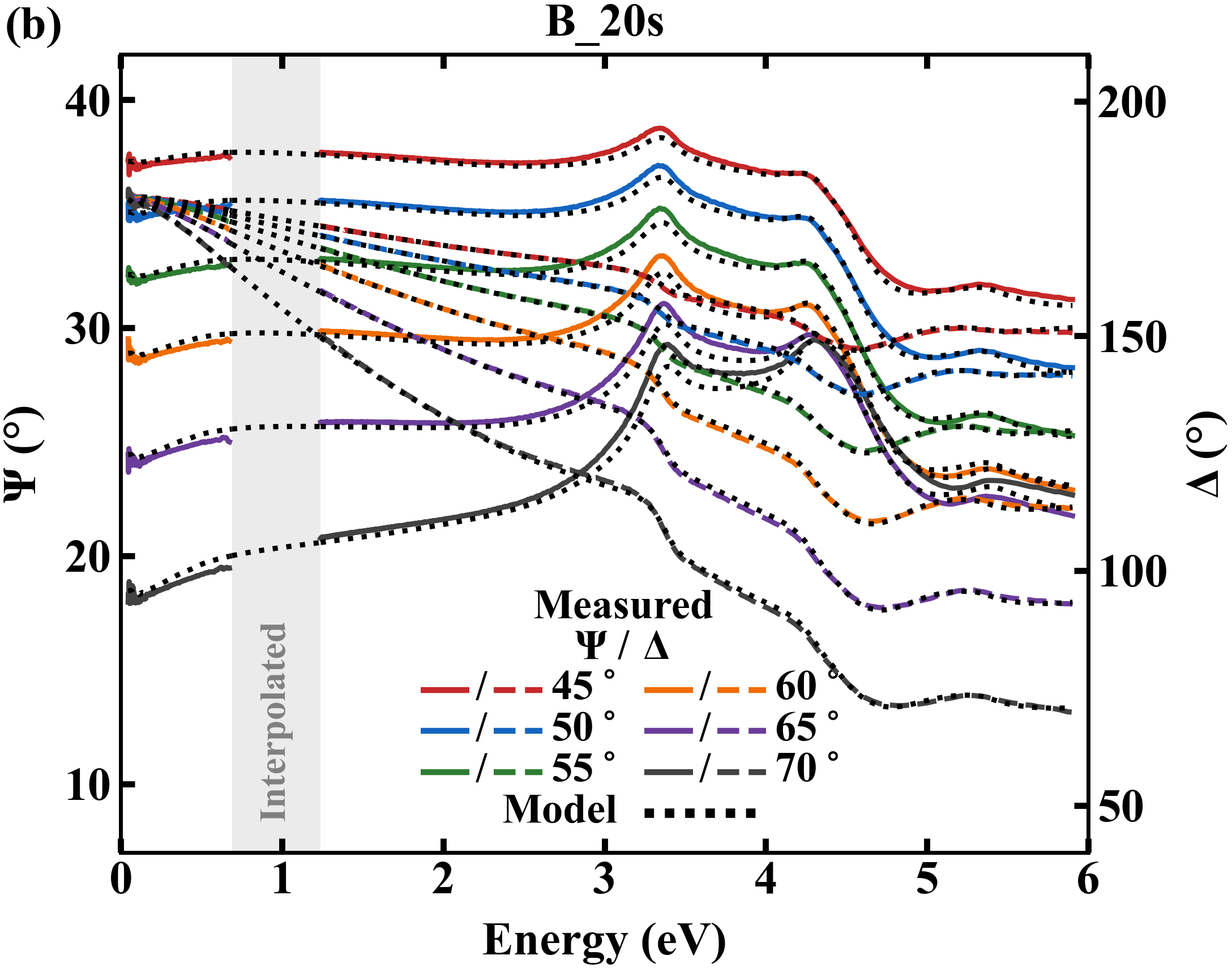} 
				\end{minipage} 
				
				\vspace{0.25cm} 
				
				\begin{minipage}{0.490\textwidth} 
					\centering 
					\includegraphics[width=\linewidth]{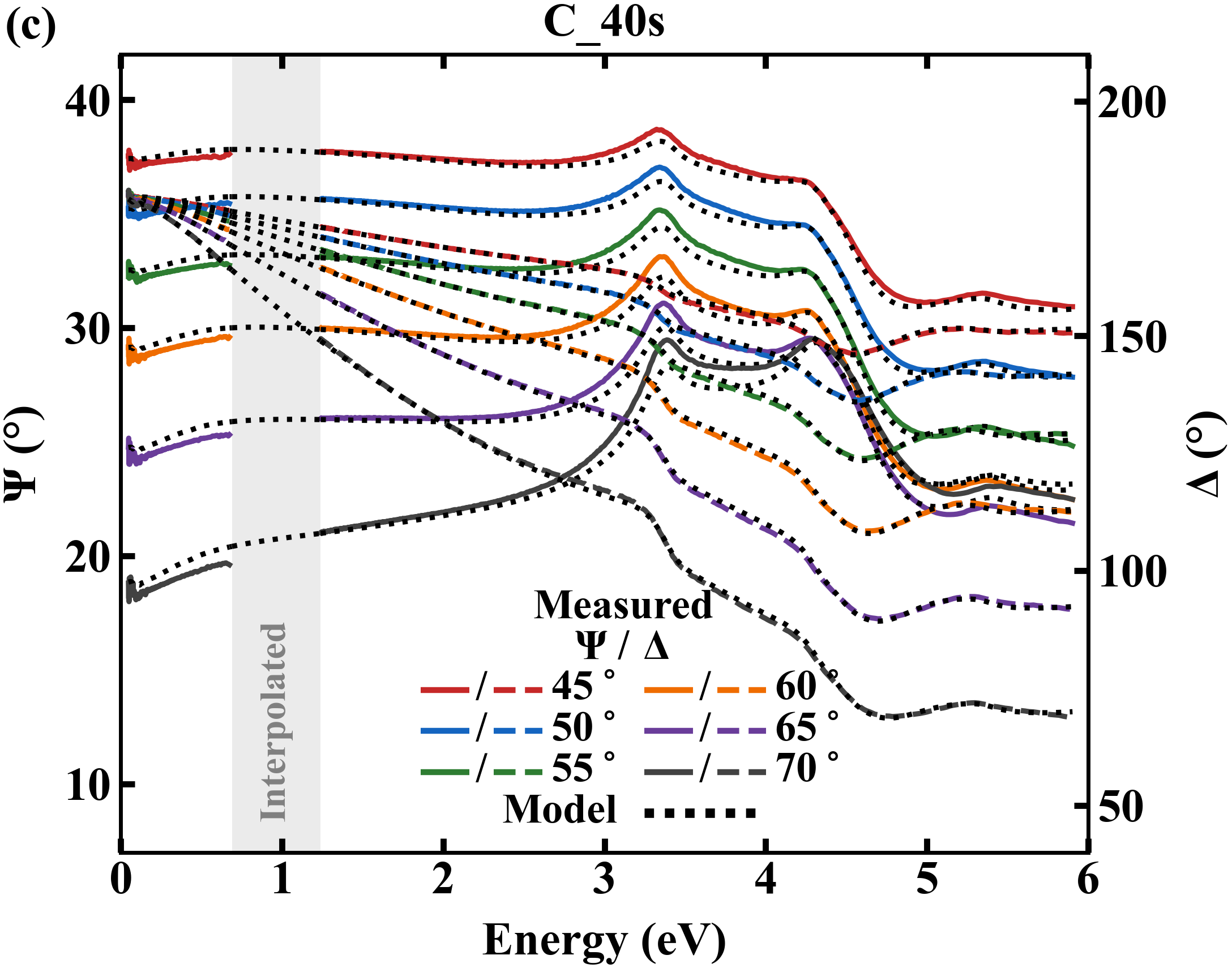} 
				\end{minipage}\hfill 
				\begin{minipage}{0.490\textwidth} 
					\centering 
					\includegraphics[width=\linewidth]{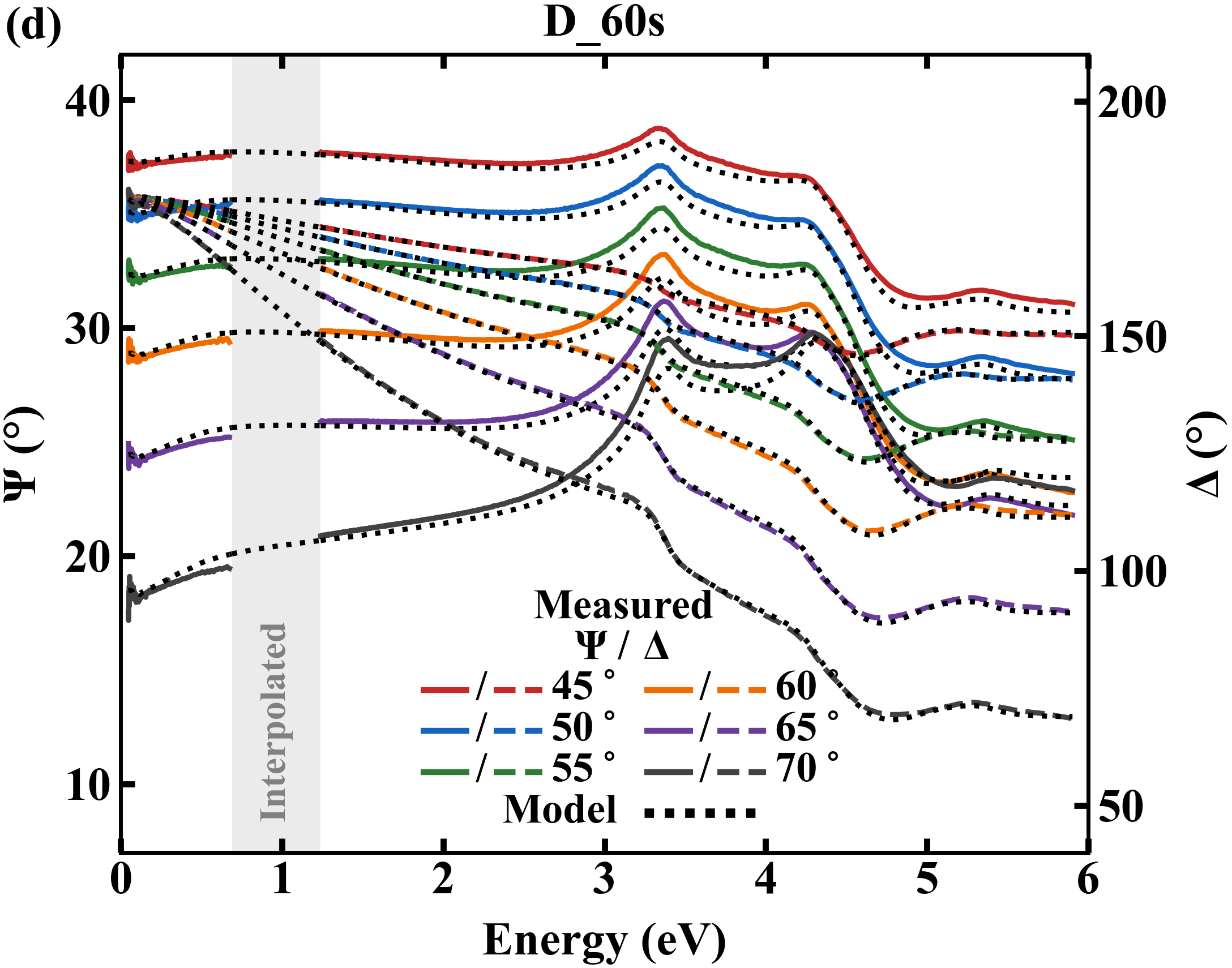} 
				\end{minipage}
				
			\end{minipage}%
		}
		
		\caption{Measured and modeled ellipsometric $\Psi$ and $\Delta$ spectra of NbN thin films after oxygen plasma treatment for (a) 0\,s, (b) 20\,s, (c) 40\,s, and (d) 60\,s, acquired at incidence angles from $45^\circ$ to $70^\circ$ in $5^\circ$ steps. Colored solid and dashed curves represent the measured $\Psi$ and $\Delta$ spectra, while the black dotted curves show the model calculations. The region between 0.82 and 1.24\,eV, shaded in gray, was obtained by interpolation within the optical model while maintaining Kramers--Kronig consistency.} 
		\label{fig:psi_delta_after} 
	\end{figure*}
	
	\begin{table*}
		\caption{Drude--Lorentz model parameters obtained from spectroscopic ellipsometry analysis of NbN thin films after oxygen plasma treatment for 0, 20, 40, and 60\,s. Here, Amp., Br., and En. denote the oscillator amplitude, broadening, and resonance energy, respectively, while $N$, $\upmu$, $m^*$, and $\upvarepsilon_{\infty}$ represent the charge-carrier density, mobility, effective mass, and high-frequency dielectric constant.}
		\label{tab:oscillator_parameters_after}
		\begin{ruledtabular}
			\begin{tabular*}{\textwidth}{@{\extracolsep{\fill}}lcccccccccccccc@{}}
				
				& \multicolumn{3}{c}{1\textsuperscript{st} UV Lorentz}
				& \multicolumn{3}{c}{2\textsuperscript{nd} UV Lorentz}
				& \multicolumn{3}{c}{IR Lorentz}
				& \multicolumn{3}{c}{Drude}
				& & \\
				
				Sample
				& Amp. & Br. & En.
				& Amp. & Br. & En.
				& Amp. & Br. & En.
				& $N$ & $\upmu$ & $m^*$
				& $\upvarepsilon_{\infty}$ & Roughness \\
				
				& & (eV) & (eV)
				& & (eV) & (eV)
				& & (eV) & (eV)
				& ($10^{22}$\,cm$^{-3}$)
				& (cm$^2$\,V$^{-1}$\,s$^{-1}$)
				&
				&
				& (nm) \\
				
				\hline
				
				A\_0s
				& 6.64 & 7.54 & 7.33
				& --- & --- & ---
				& 43.00 & 2.37 & 0.81
				& 1.20
				& 2.44
				& 1.82
				& 1.00
				& 1.45\\
				
				B\_20s
				& 5.73 & 4.11 & 6.67
				& 2.31 & 1.39 & 4.94
				& 38.00 & 2.45 & 0.83
				& 1.08
				& 2.41
				& 1.80
				& 2.39
				& 1.76 \\
				
				C\_40s
				& 5.22 & 3.92 & 6.44
				& 1.92 & 1.31 & 4.92
				& 36.10 & 2.45 & 0.83
				& 1.05
				& 2.36
				& 1.80
				& 2.70
				& 1.39 \\
				
				D\_60s
				& 5.13 & 4.65 & 6.40
				& 1.27 & 1.12 & 4.95
				& 34.00 & 2.35 & 0.83
				& 1.03
				& 2.27
				& 1.80
				& 2.55
				& 1.09 \\
				
			\end{tabular*}
		\end{ruledtabular}
	\end{table*}
	
	The Drude--Lorentz dispersion model consists of three Lorentz oscillators and one Drude term. The Lorentz oscillators describe the bound-electron contributions, whereas the Drude term accounts for the free-carrier response \cite{Silaeva2021}. The contribution of each Lorentz oscillator to the dielectric function is calculated using the relation in Eq.~\ref{eq:lorentz}, taken from Ref.~\cite{JAWoollam2020}:
	\begin{equation}
		\upvarepsilon_{\mathrm{Lorentz}}(E)=
		\frac{\mathrm{Amp}\cdot\mathrm{Br}\cdot\mathrm{En}}
		{\mathrm{En}^{2}-E^{2}-iE\mathrm{Br}},
		\label{eq:lorentz}
	\end{equation}
	where Amp is the oscillator amplitude, Br is the broadening parameter, En is the resonance energy, $E$ is the photon energy, and $i$ is the imaginary unit. The Drude contribution, as described in Ref.~\cite{JAWoollam2020}, is similarly expressed by Eq.~\ref{eq:drude}:
	
	\begin{equation}
		\upvarepsilon_{\mathrm{Drude}(N\upmu)}(E)=
		\frac{-\hbar^{2}q^{2}N\upmu}
		{\upvarepsilon_{0}\left(\upmu m_{e}m^{*}\cdot E^{2}+iq\hbar E\right)},
		\label{eq:drude}
	\end{equation}	
	where $N$ is the carrier concentration, $q$ is the elementary charge, $m^{*}$ is the dimensionless effective mass, $\upmu$ is the carrier mobility, $m_{e}$ is the electron mass, $\upvarepsilon_{0}$ is the vacuum permittivity, and $\hbar$ is the reduced Planck constant.

	The optimized model parameters are summarized in Table~\ref{tab:oscillator_parameters_after}. The physical significance of the individual oscillators and their evolution with oxygen plasma treatment will be discussed in Sec.~\ref{sec:discussion}. Besides the Drude and Lorentz oscillator parameters, the high-frequency dielectric constant ($\upvarepsilon_{\infty}$) and the surface roughness were also obtained from the fitting procedure. The extracted $\upvarepsilon_{\infty}$ values were 1.0, 2.4, 2.7, and 2.6 for A\_0s, B\_20s, C\_40s, and D\_60s, respectively. Since $\upvarepsilon_{\infty}$ represents the contribution of electronic transitions outside the investigated spectral range \cite{JAWoollam2020}, the observed increase after oxygen plasma treatment indicates changes in the high-energy dielectric response. The surface roughness obtained using the effective-medium approximation (EMA) was 1.5\,nm for A\_0s, 1.8\,nm for B\_20s, 1.4\,nm for C\_40s, and 1.1\,nm for D\_60s.
	
	\subsection{Optical properties}
	
	Based on the fitted Drude--Lorentz model, the complex dielectric function of the NbN layer was obtained by combining the high-frequency dielectric constant ($\upvarepsilon_{\infty}$), three Lorentz oscillators, and the Drude contribution, as expressed in Eq.~\ref{eq:eps_total}:
	
	\begin{equation}
		\begin{split}
			\upvarepsilon_{\mathrm{Total}}(E)=\,
			&\upvarepsilon_{\infty}
			+\upvarepsilon_{\mathrm{UV-Lorentz\,1}}(E)
			+\upvarepsilon_{\mathrm{UV-Lorentz\,2}}(E) \\
			&+\upvarepsilon_{\mathrm{IR-Lorentz}}(E)
			+\upvarepsilon_{\mathrm{Drude}(N\upmu)}(E).
		\end{split}
		\label{eq:eps_total}
	\end{equation}

	The refractive index ($n$) and extinction coefficient ($k$) were calculated from the complex dielectric function using Eq.~\ref{eq:nk}, according to Ref.~\cite{Xu2018}:
	
	\begin{equation}
		n(E)=
		\sqrt{\frac{|\upvarepsilon|+\upvarepsilon_1}{2}},
		\qquad
		k(E)=
		\sqrt{\frac{|\upvarepsilon|-\upvarepsilon_1}{2}},
		\label{eq:nk}
	\end{equation}
	where $|\upvarepsilon|=\sqrt{\upvarepsilon_1^2+\upvarepsilon_2^2}$. The reflectivity of the NbN layer, including the oxygen-modified surface region, was calculated from the extracted $n$ and $k$ values using the normal-incidence Fresnel relation \cite{Stenzel2026}, as expressed in Eq.~\ref{eq:reflectivity}, and is denoted as $R_{\mathrm{eff}}$:
	
	\begin{equation}
		R_{\mathrm{eff}}(E)=
		\frac{[n(E)-1]^2+k(E)^2}
		{[n(E)+1]^2+k(E)^2}.
		\label{eq:reflectivity}
	\end{equation}
	
\begin{figure}[t] 
		\centering 
		\nolinenumbers 
		
		\includegraphics[width=0.77\linewidth]{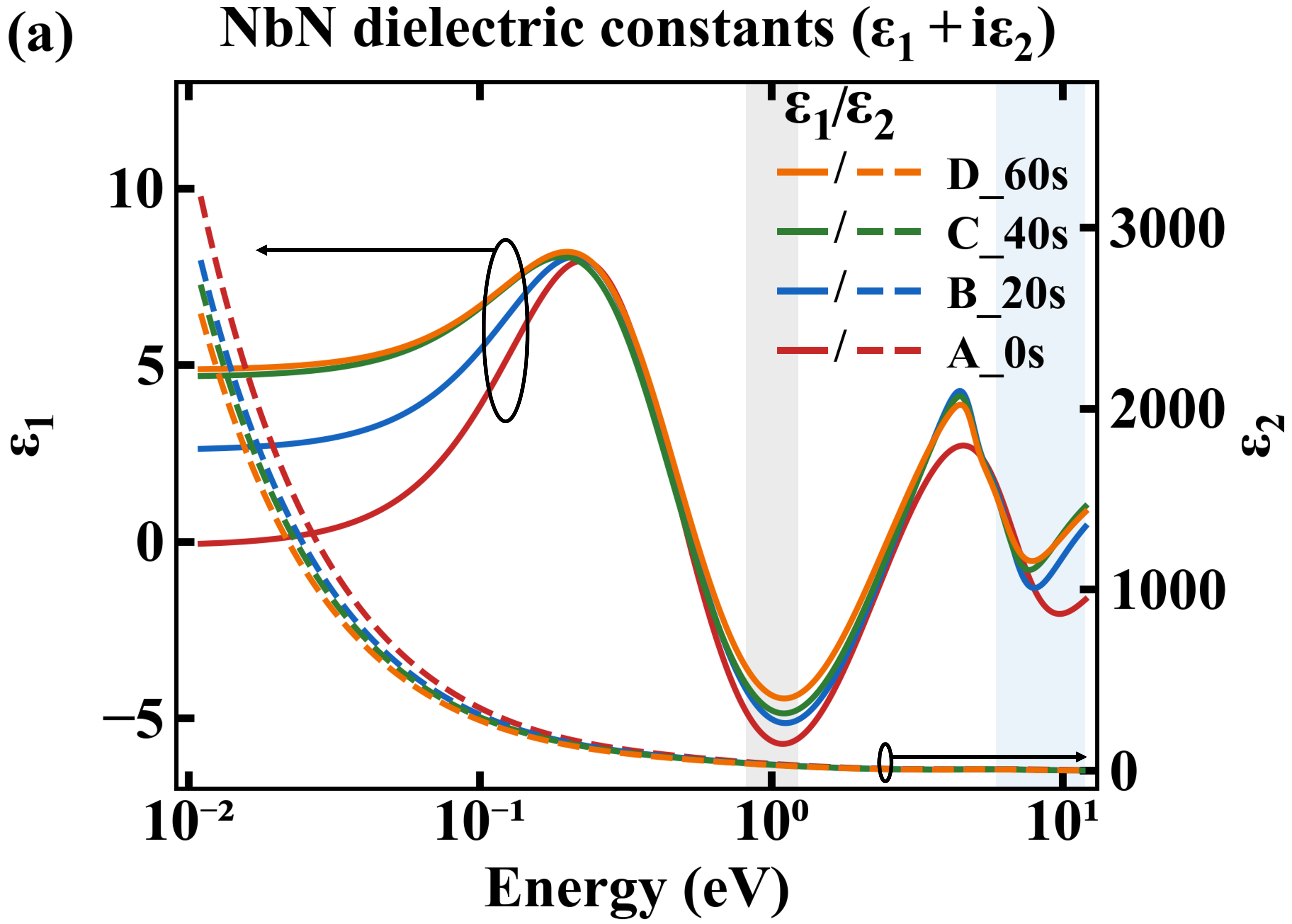} 
		
		\vspace{0.1cm} 
		
		\includegraphics[width=0.77\linewidth]{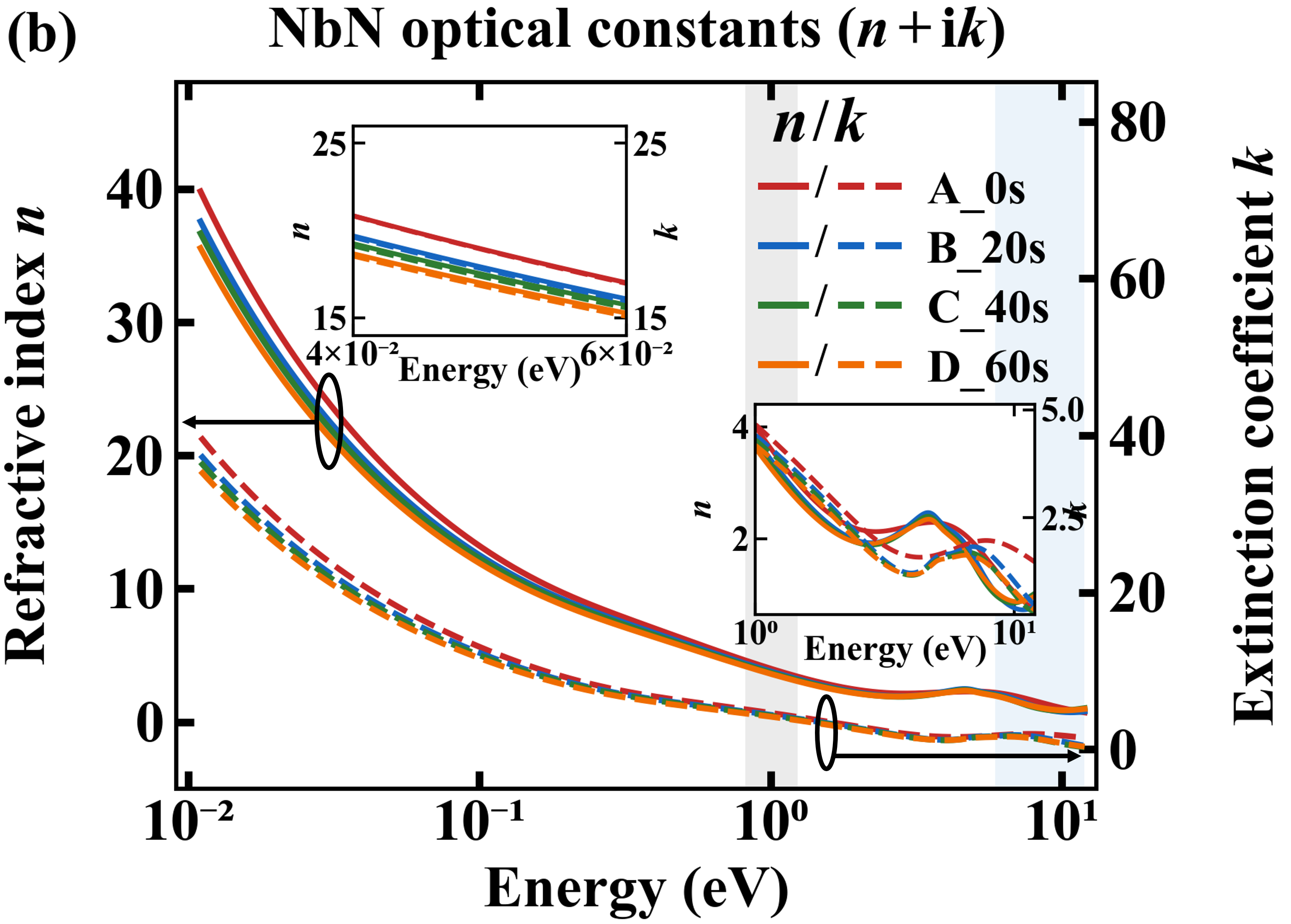} 
		
		\vspace{0.1cm}
		
		\includegraphics[width=0.77\linewidth]{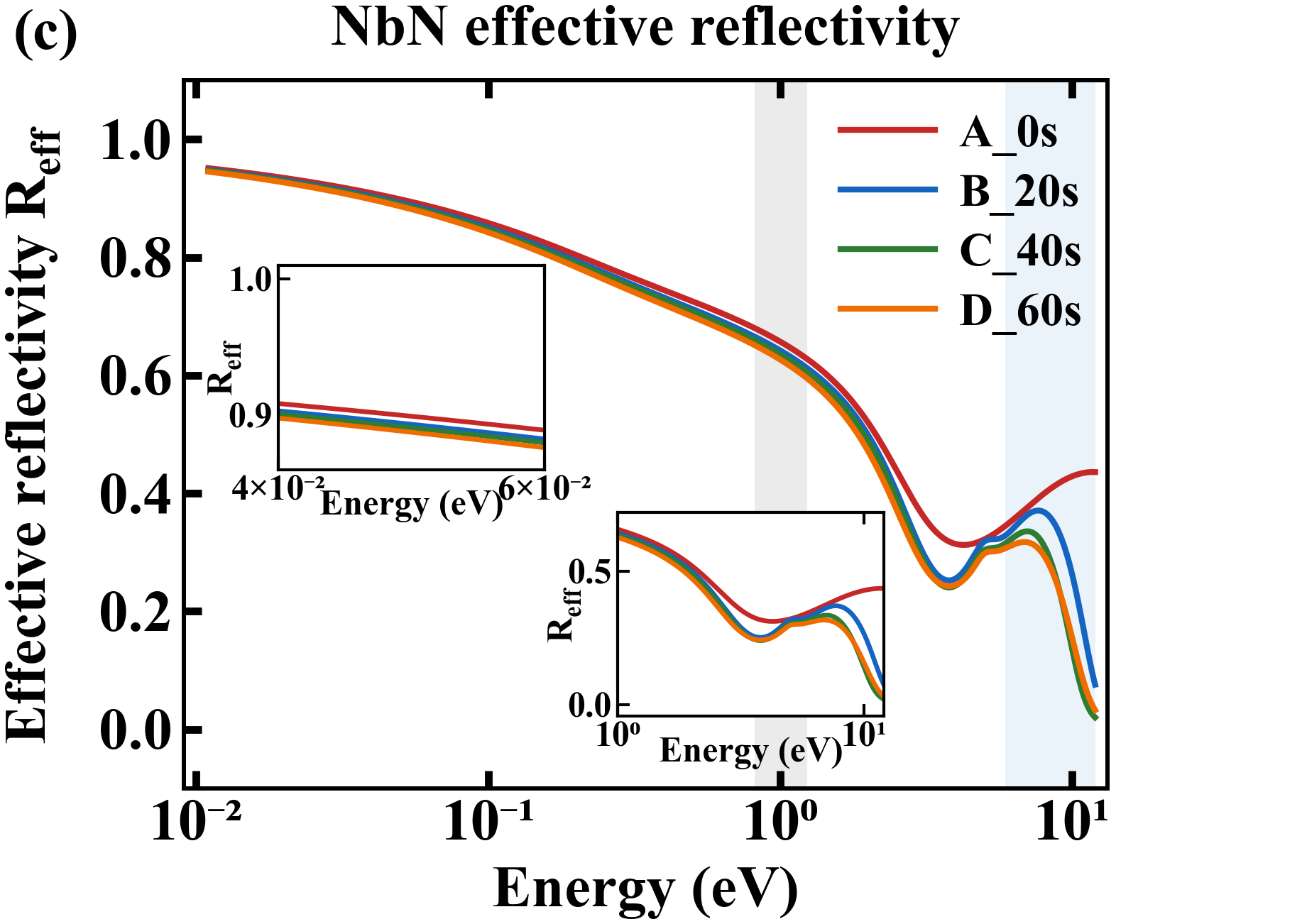}
		\caption{(a) $\upvarepsilon_1$ and $\upvarepsilon_2$ parts of the complex dielectric function, (b) $n$ and $k$, and (c) effective reflectivity $R_{\mathrm{eff}}$ of NbN thin films after oxygen plasma treatment for 0, 20, 40, and 60\,s. The dielectric functions and optical constants were extracted by fitting the spectroscopic ellipsometry data with the Drude--Lorentz dielectric-function model, and $R_{\mathrm{eff}}$ was calculated from the resulting $n$ and $k$ values. The insets in panels (b) and (c) show magnified views of the low- and high-energy spectral regions, respectively, allowing a clearer comparison of the spectral changes induced by oxygen plasma treatment. The gray- and blue-shaded regions indicate the interpolated (0.82--1.24\,eV) and extrapolated (6.20--12\,eV) responses, respectively, both described by the same Kramers--Kronig-consistent optical model.}
		\label{fig:dielectric_optical_after} 
	\end{figure}

	The real and imaginary parts of the complex dielectric function, the corresponding optical constants, and the calculated effective reflectivity are presented in Fig.~\ref{fig:dielectric_optical_after}. The real part, $\upvarepsilon_1$, describes the dispersive response of the material, whereas the imaginary part, $\upvarepsilon_2$, is associated with optical absorption \cite{Subramanian2026,Castro1999}. The refractive index $n$ and extinction coefficient $k$ characterize the propagation and attenuation of light within the NbN films, respectively \cite{Speets2024,Makhlouka2022}.

	A notable feature of the dielectric response is the evolution of $\upvarepsilon_1$ in the low-energy region (see Fig.~\hyperref[fig:dielectric_optical_after]{\ref*{fig:dielectric_optical_after}(a)}). While the reference sample A\_0s exhibits a slightly negative $\upvarepsilon_1$, with a magnitude considerably smaller than typically expected for a good metal \cite{Ordal1983}, $\upvarepsilon_1$ becomes positive after oxygen plasma treatment. At the same time, the increase in $\upvarepsilon_2$ toward lower photon energies indicates that the free-carrier response remains significant in all samples. This behavior can be understood from the individual contributions of the Drude--Lorentz model as shown in Fig.~\ref{fig:oscillators_after}. The relatively strong carrier scattering in these NbN films limits the magnitude of the negative Drude contribution to $\upvarepsilon_1$ at low photon energies \cite{Park2023}. Consequently, the negative Drude contribution is increasingly outweighed by the positive dispersive contributions of the Lorentz terms, particularly the IR Lorentz oscillator in the low-energy region. The resulting total $\upvarepsilon_1$ therefore shifts from slightly negative for A\_0s to positive after oxygen plasma treatment. Thus, the positive $\upvarepsilon_1$ is possibly due to the surface oxidation of the NbN films \cite{Yerra2026} and the combined contributions to the dielectric function. The optical constants, $n$ and $k$, decrease with increasing oxygen plasma treatment time at low photon energies (Fig.~\hyperref[fig:dielectric_optical_after]{\ref*{fig:dielectric_optical_after}(b)}), although the differences among the samples remain small. Both exhibit a strong decrease with increasing photon energy, followed by changes in their spectral profiles at higher energies. The treatment-dependent differences become more evident in $k$ toward the ultraviolet range, consistent with the modifications in the dielectric response following oxygen plasma treatment. At low photon energies, the plasma-treated samples exhibit slightly lower reflectivity, $R_{\mathrm{eff}}$, than the reference sample, as illustrated in Fig.~\hyperref[fig:dielectric_optical_after]{\ref*{fig:dielectric_optical_after}(c)}. This reduction becomes more evident at higher energies, particularly in the extrapolated region above approximately 7\,eV, where $R_{\mathrm{eff}}$ generally decreases with increasing oxygen plasma treatment time.

	\section{Discussion}
	\label{sec:discussion}
	
	\begin{figure*}[t] 
		\centering 
		\nolinenumbers 
		
		\resizebox{0.95\textwidth}{!}{%
			\begin{minipage}{\textwidth}
				\centering
				
				\begin{minipage}{0.490\textwidth} 
					\centering 
					\includegraphics[width=\linewidth]{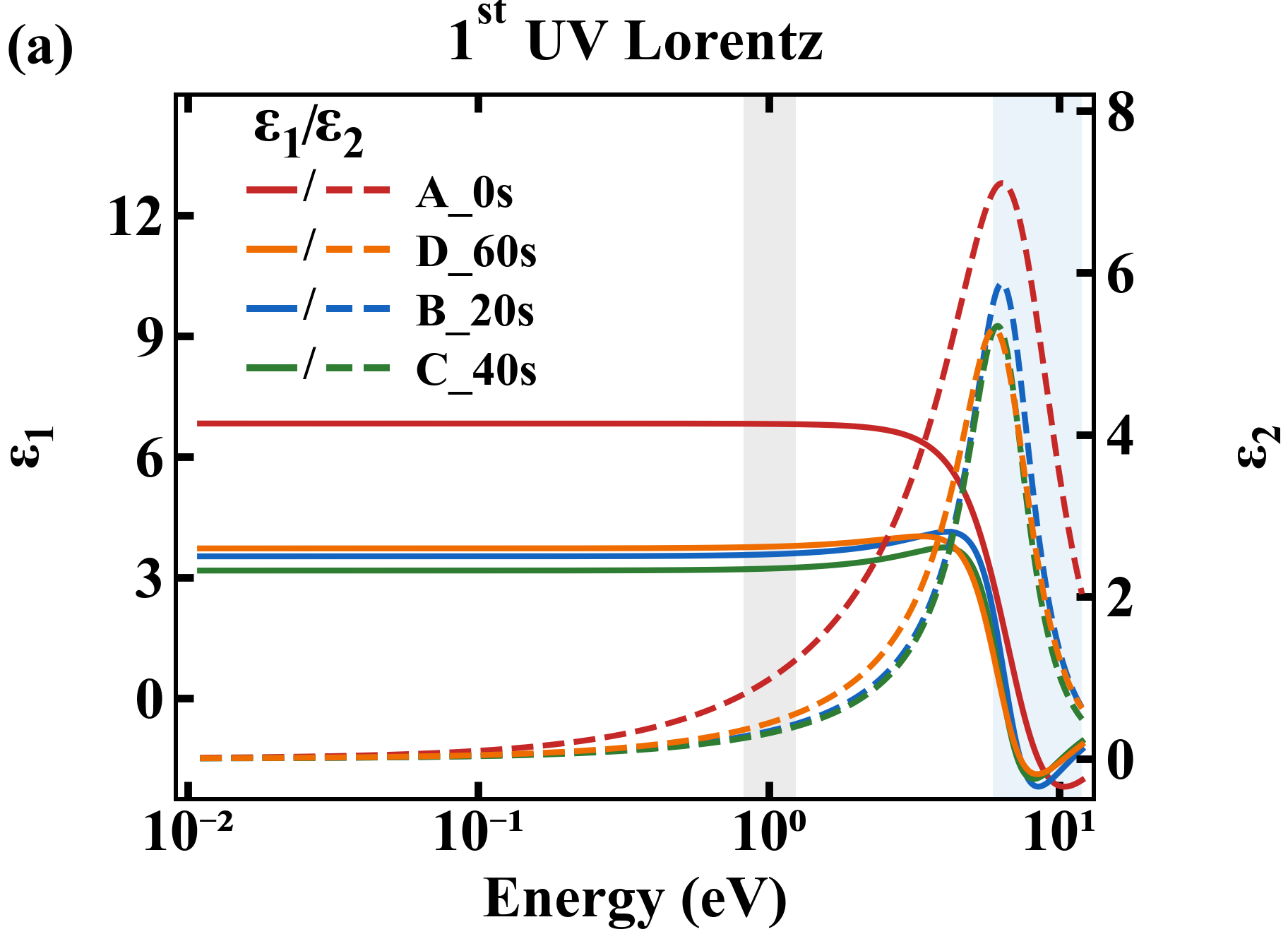} 
				\end{minipage}\hfill 
				\begin{minipage}{0.490\textwidth} 
					\centering 
					\includegraphics[width=\linewidth]{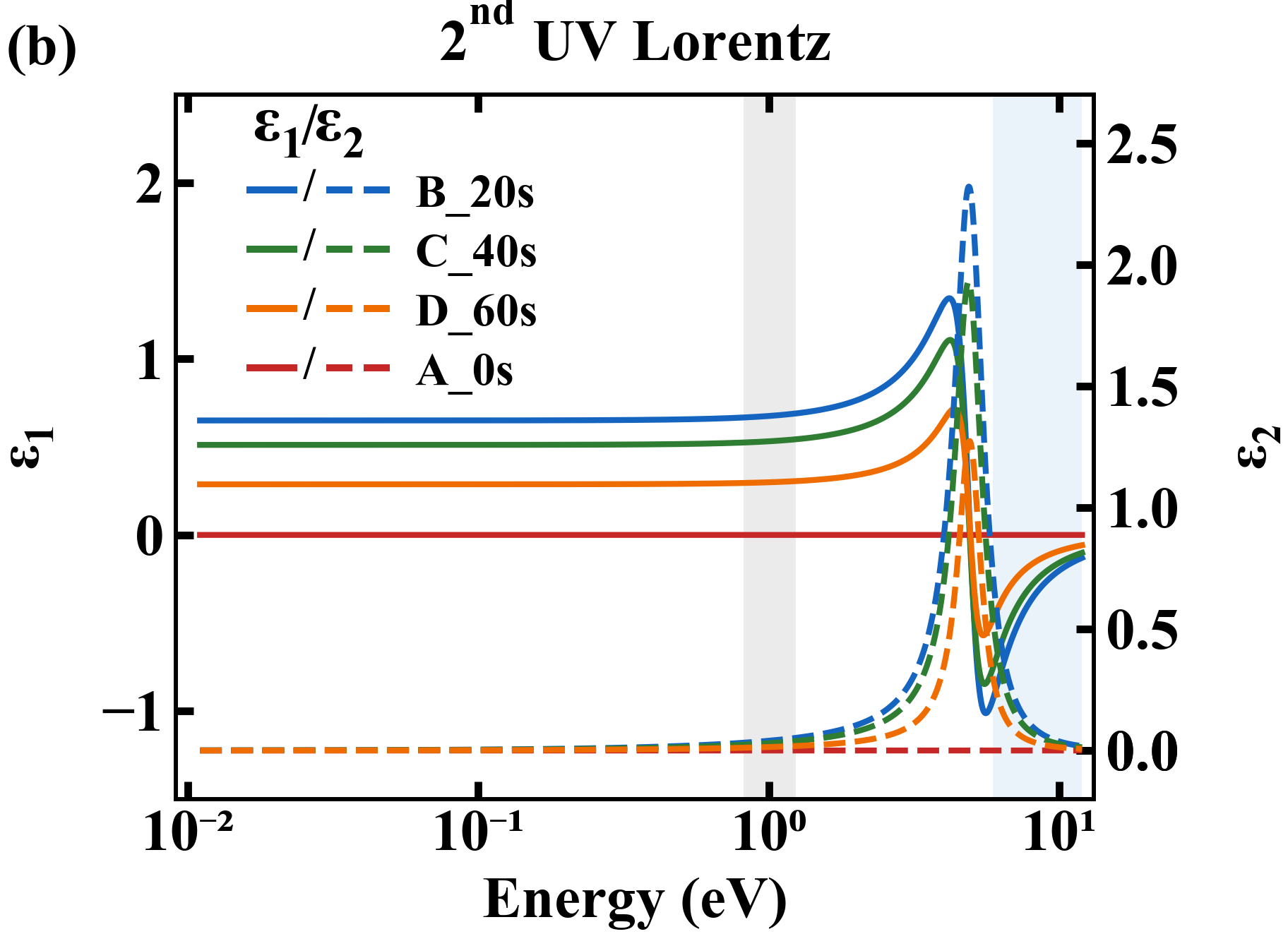} 
				\end{minipage} 
				
				\vspace{0.25cm} 
				
				\begin{minipage}{0.490\textwidth} 
					\centering 
					\includegraphics[width=\linewidth]{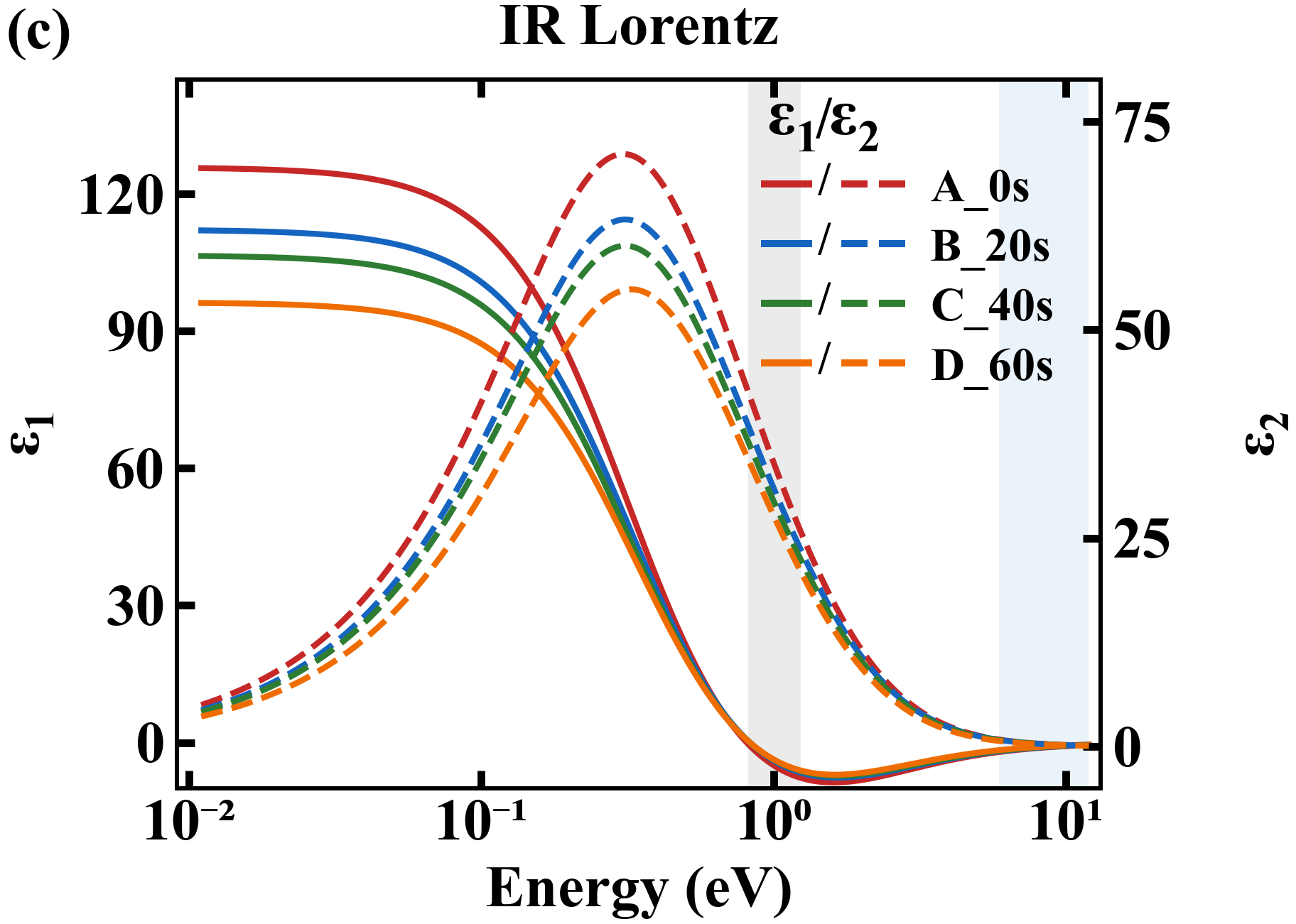} 
				\end{minipage}\hfill 
				\begin{minipage}{0.490\textwidth} 
					\centering 
					\includegraphics[width=\linewidth]{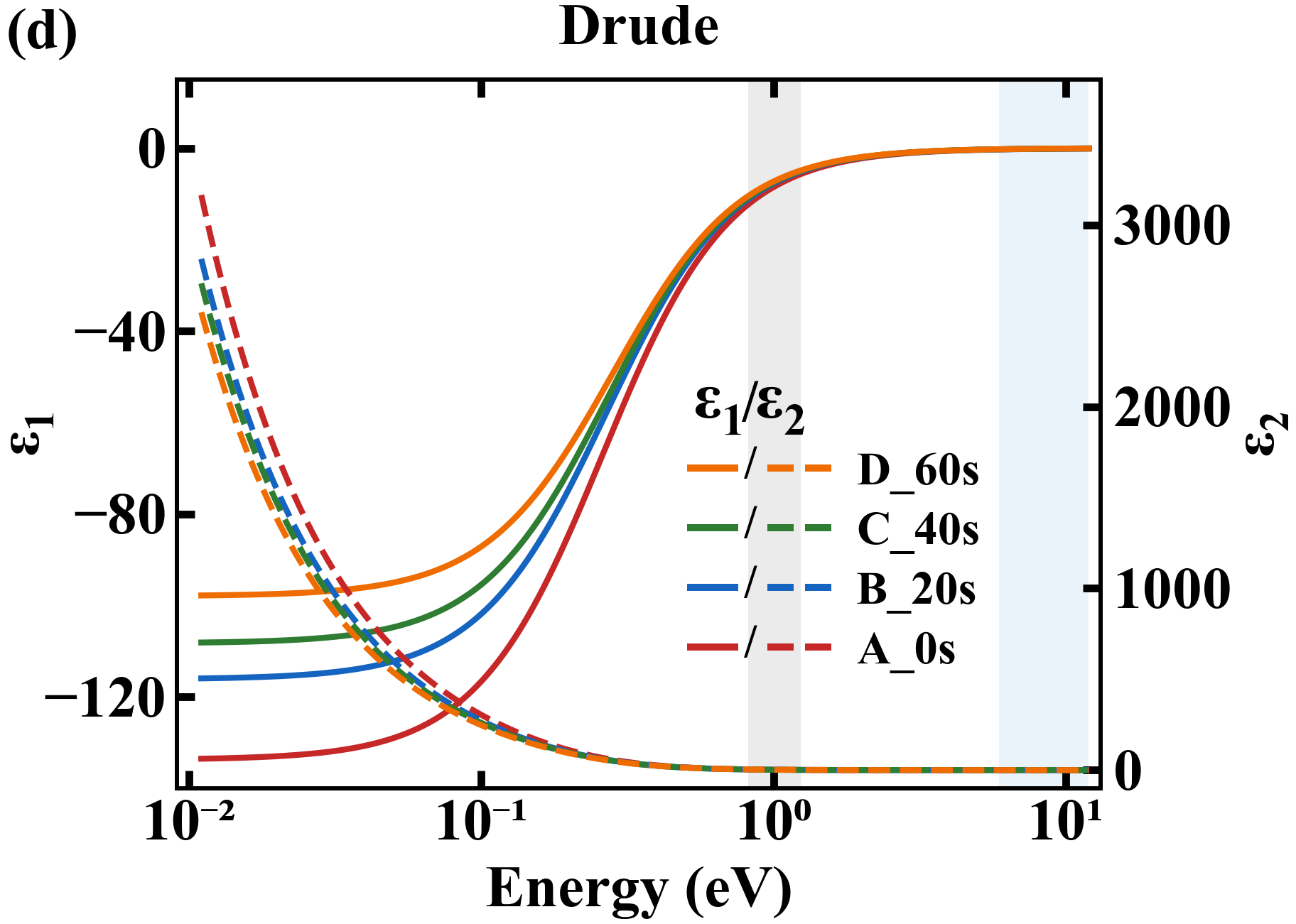} 
				\end{minipage} 
				
			\end{minipage}%
		}
		
		\caption{Dielectric-function contributions of the Drude--Lorentz model components for NbN films after oxygen plasma treatment for 0, 20, 40, and 60\,s: (a) the $1^{\mathrm{st}}$ UV Lorentz oscillator, describing the dominant interband absorption involving hybridized Nb 4$d$ and N 2$p$ states; (b) the $2^{\mathrm{nd}}$ UV Lorentz oscillator, associated with oxygen-related electronic transitions in the oxidized NbO$_x$ surface region; (c) the IR Lorentz oscillator, describing the low-energy anti-Drude response; and (d) the Drude term, representing the free-carrier response. The gray-shaded region (0.82--1.24\,eV) indicates the interpolated response, while the blue-shaded region (6.20--12\,eV) corresponds to the extrapolated response, with both regions maintaining Kramers--Kronig consistency within the same optical model.} 
		\label{fig:oscillators_after} 
	\end{figure*}

	The dielectric function and optical constants presented in the previous section represent the combined response of the Drude term and the Lorentz oscillators. The individual oscillator contributions were analyzed separately to clarify their physical origin and evolution after oxygen plasma treatment.
	
	The $1^{\mathrm{st}}$ UV Lorentz oscillator, plotted in
	Fig.~\hyperref[fig:oscillators_after]{\ref*{fig:oscillators_after}(a)}, represents the dominant
	ultraviolet interband absorption of the NbN films. Its spectral position around 7\,eV is consistent with the high-energy
	interband feature previously reported for NbN and is attributed to electronic transitions involving strongly hybridized Nb 4$d$ and N 2$p$ states \cite{Kern2024, Farha2018, Sanjines2006}. Following oxygen plasma treatment, the oscillator energy shifts from 7.33\,eV for A\_0s to 6.67, 6.44, and 6.40\,eV for B\_20s, C\_40s, and D\_60s, respectively (see Table~\ref{tab:oscillator_parameters_after}). This red shift is attributed to changes in the electronic structure associated with the formation of Nb--O bonds, as oxidation of NbN promotes niobium-oxide formation and modifies the electronic states involved in optical transitions \cite{Ermolieff1986, Hossain2025, Oki2016}. Meanwhile, the decrease in oscillator amplitude is consistent with a redistribution of spectral weight associated with changes in the NbN interband transitions and the emergence of oxygen-related electronic transitions \cite{ElShazly2018, Fenker2006}, while the reduced broadening corresponds to a narrower ultraviolet absorption band.

	The emergence of the $2^{\mathrm{nd}}$ UV Lorentz oscillator after oxygen plasma treatment constitutes one of the key findings of the present study. As shown in Fig.~\hyperref[fig:oscillators_after]{\ref*{fig:oscillators_after}(b)}, this absorption is not observed in A\_0s and appears only after oxygen plasma treatment in B\_20s, C\_40s, and D\_60s. The absorption peak remains nearly constant around 4.9\,eV, whereas the oscillator amplitude decreases from 2.31 for B\_20s to 1.92 for C\_40s and 1.27 for D\_60s (see Table~\ref{tab:oscillator_parameters_after}). Previous studies have demonstrated that oxidation promotes the formation of NbO$_x$, particularly Nb$_2$O$_5$, on the NbN surface \cite{Ermolieff1985,Lubenchenko2017,Jouve1996,Frankenthal1983}. The emergence of the additional UV absorption observed only after oxygen plasma treatment is therefore associated with electronic transitions originating from this oxygen-rich surface region. Based on this interpretation, the UV absorption of the plasma-treated samples can no longer be described solely by the original NbN UV interband transition represented by the $1^{\mathrm{st}}$ UV Lorentz oscillator. Instead, an additional Lorentz oscillator is required to account for the optical response introduced by the oxidized surface layer.
	
	In contrast to the UV Lorentz oscillators, which arise from interband electronic transitions, the IR Lorentz oscillator, displayed in Fig.~\hyperref[fig:oscillators_after]{\ref*{fig:oscillators_after}(c)}, describes the anti-Drude (also referred to as an anomalous or displaced Drude peak) absorption contribution in the infrared region \cite{Kern2024, Fratini2021, Kern2021}, which originates from disorder and deviations from ideal free-carrier behavior in metallic NbN \cite{Kern2024, Fratini2021}. The resonance energy remains nearly unchanged at approximately 0.82\,eV for all samples, while its amplitude decreases with increasing oxygen plasma treatment time. The preserved oscillator energy indicates that oxygen plasma treatment does not alter the fundamental origin of the anti-Drude absorption, whereas oxygen-induced disorder and enhanced carrier scattering reduce its contribution to the infrared optical response.
	
	\begin{figure}[t]
		\centering
		\includegraphics[width=\columnwidth]{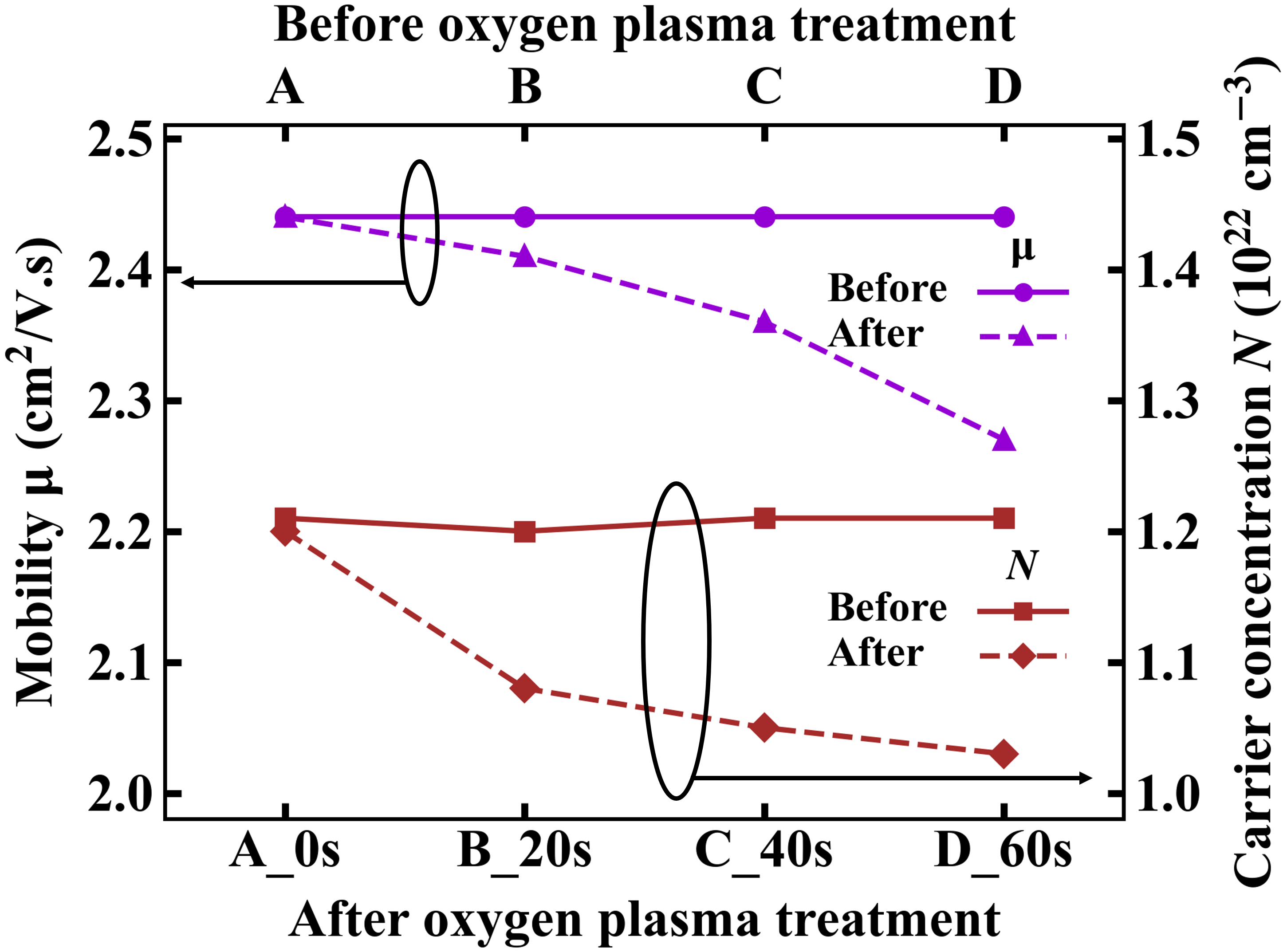}
		\caption{Variation of carrier concentration ($N$) and mobility ($\upmu$) of NbN thin films as a function of oxygen plasma treatment time. Both parameters show a gradual decrease with increasing treatment duration, indicating that oxygen incorporation during plasma treatment leads to enhanced carrier scattering and a reduction in free-carrier density.}
		\label{fig:mobility_carrier_concentration}
	\end{figure}

	The low-energy optical response of the NbN films is dominated by the Drude contribution, as presented in Fig.~\hyperref[fig:oscillators_after]{\ref*{fig:oscillators_after}(d)}. The Drude term accounts for the optical behavior of free carriers \cite{Schubert2004,Fomra2024}. After oxygen plasma treatment, the magnitudes of both the $\upvarepsilon_{1}$ and $\upvarepsilon_{2}$ parts of the Drude contribution decrease at low photon energies, reflecting a reduction in the free-carrier optical response. Within the Drude model, this response is characterized by the effective carrier concentration ($N$) and effective carrier mobility ($\upmu$). The $N$ decreases from $1.20 \times 10^{22}$\,cm$^{-3}$ for A\_0s to $1.03 \times 10^{22}$\,cm$^{-3}$ for D\_60s, while the $\upmu$ decreases from 2.44 to 2.27\,cm$^{2}$\,V$^{-1}$\,s$^{-1}$ with increasing oxygen plasma treatment time (see Fig.~\ref{fig:mobility_carrier_concentration} and Table~\ref{tab:oscillator_parameters_after}). We propose that the reduction in $N$ originates from the partial depletion of delocalized electrons accompanying Nb--O bond formation. The decrease in $\upmu$ may result from enhanced carrier scattering associated with oxygen-induced disorder. In addition, the formation of an NbO$_x$ surface layer reduces the effective thickness of the conductive NbN layer and may therefore contribute to the decrease in the Drude-derived mobility.

	While $N$ and $\upmu$ obtained from the Drude analysis characterize the free-carrier response of the NbN films, these quantities were further used to derive additional transport parameters and thereby evaluate the influence of oxygen plasma treatment on their electronic transport behavior. Using the extracted values of $N$, $\upmu$, and the dimensionless effective mass $m^{*}$, the plasma frequency ($\upomega_{p}$), direct-current (DC) conductivity ($\upsigma$), electrical resistivity ($\uprho$), carrier scattering time ($\uptau$), and scattering rate ($\Gamma$) were calculated. The plasma frequency was determined from $\upomega_{p}=\sqrt{Ne^{2}/(\upvarepsilon_{0}m)}$, where $e$ is the elementary charge, $\upvarepsilon_{0}$ is the vacuum permittivity, and $m=m_{e}m^{*}$ is the effective carrier mass, with $m_{e}$ denoting the free-electron mass \cite{Chochol2017}. Within the Drude framework, $\upomega_{p}$ depends on both the carrier concentration and effective mass and defines the characteristic frequency at which the conduction electrons collectively oscillate in response to a displacement from their equilibrium position \cite{Cunha2020}. The DC conductivity was evaluated from $\upsigma=Ne\upmu$, which relates electrical conduction to the concentration and mobility of the available charge carriers \cite{Smith2001}. The associated resistivity was obtained from $\uprho=1/\upsigma$ and represents the opposition of the material to charge transport \cite{Pandian2022}. Accordingly, variations in $\upsigma$ and $\uprho$ provide complementary measures of how oxygen plasma treatment influences electrical conduction in the NbN films. 
	The carrier scattering time was calculated using the Drude relation $\uptau=\upmu m/e$, where $\upmu$ is the carrier mobility and $m$ is the effective carrier mass, and the scattering rate was obtained from its reciprocal, $\Gamma=1/\uptau$ \cite{Zhang2021}. Within the Drude model, $\uptau$ characterizes the average timescale for momentum relaxation, while $\Gamma$ describes the frequency of momentum-relaxing scattering processes \cite{Mitra2025}. The resulting optical transport parameters are summarized in Table~\ref{tab:drude_derived}, providing a quantitative basis for evaluating the effects of oxygen plasma treatment on the carrier density, electrical conduction, and scattering dynamics of the NbN films.

	\begin{table}[t]
	\centering
	\caption{Drude-derived optical transport parameters of the NbN films after oxygen plasma treatment.}
	\label{tab:drude_derived}
	
	\renewcommand{\arraystretch}{1.2}
	
	\begin{tabular*}{\columnwidth}{@{\extracolsep{\fill}}lccccc@{}}
		\hline\hline
		
		Sample
		& $\upomega_p$
		& $\upsigma$
		& $\uprho$
		& $\uptau$
		& $\Gamma$ \\
		
		& ($10^{15}$\,rad\,s$^{-1}$)
		& ($10^{5}$\,S\,m$^{-1}$)
		& ($\upmu\Omega\cdot$cm)
		& (fs)
		& ($10^{14}$\,s$^{-1}$) \\
		
		\hline
		
		A\_0s  & 4.58 & 4.69 & 213.2 & 2.52 & 3.96 \\
		B\_20s & 4.37 & 4.17 & 239.8 & 2.47 & 4.05 \\
		C\_40s & 4.31 & 3.97 & 251.9 & 2.42 & 4.14 \\
		D\_60s & 4.27 & 3.75 & 266.9 & 2.32 & 4.30 \\
		
		\hline\hline
	\end{tabular*}
	\end{table}
	
	\newpage
	
	\section{Conclusion and Outlook}
	
	In conclusion, the influence of oxygen plasma treatment on the optical properties of ALD-grown NbN thin films was investigated using spectroscopic ellipsometry and a Drude--Lorentz dielectric-function model. The results reveal modifications in both the interband transitions and free-carrier response of NbN. The principal finding is the emergence of an additional ultraviolet absorption band near 4.9\,eV following oxygen plasma treatment, which is attributed to oxygen-induced electronic transitions associated with NbO$_x$ formation at the film surface. The Drude analysis reveals reductions in carrier concentration and mobility with increasing oxygen plasma treatment time, indicating changes in the free-carrier response of NbN. The decrease in carrier concentration is accompanied by a reduction in the calculated effective reflectivity, $R_{\mathrm{eff}}$, particularly at higher photon energies. These findings are consistent with the influence of carrier depletion on the optical response of conducting films. In NbN-based SNSPDs, the nanowire boundaries can exhibit normal-conducting behavior due to reduced superconductivity, which may increase optical reflectivity. Oxygen plasma treatment may therefore offer a means of modifying the free-carrier response of NbN to reduce reflectivity and potentially enhance photon absorption. These results establish a basis for further investigations of the relationship between surface oxidation, optical absorption, and superconducting performance in NbN-based SNSPDs.

	Controlled surface oxidation could provide an additional means of tuning optical coupling in NbN-based superconducting detectors. At temperatures below $T_c$ of the underlying NbN, the chemically modified surface region may remain nonsuperconducting or exhibit suppressed superconductivity \cite{Charaev2017}. Conversion of near-surface NbN into oxides can reduce the free-carrier contribution to the effective optical response and modify the dielectric contrast between the surface region and the remaining superconducting layer. These changes may reduce reflectance over a suitable spectral range and redistribute the optical field within the film. If this redistribution increases absorption specifically in the superconducting NbN, a larger fraction of incident photons would deposit energy in the active layer. For photon energies well above the superconducting gap, $\hbar\omega \gg 2\Delta(T)$, this absorbed energy generates nonequilibrium quasiparticles and initiates further Cooper-pair breaking during relaxation \cite{Beck2011}. An increase in absorption within the superconducting layer could therefore improve the optical contribution to SNSPD detection efficiency \cite{Hadfield2009}.

	Establishing this possibility requires evaluating the optical response of the complete oxide/NbN/substrate structure at the relevant photon wavelengths and operating temperatures. Reduced reflectance alone does not establish enhanced absorption in NbN, since it can also be accompanied by increased transmission or absorption in the oxide and substrate. Furthermore, the effective free-carrier concentration obtained from normal-state optical fitting should be distinguished from the superconducting condensate density. Future work should therefore combine temperature-dependent optical measurements with multilayer calculations that determine the absorption in each constituent layer. Comparison with the superconducting transition temperature, switching current, and detector efficiency would then establish whether any optical benefit is retained at the device level. Such studies could identify oxidation conditions that enhance photon absorption while preserving sufficient superconducting thickness and current-carrying capacity.		
	
	\begin{acknowledgments}
		We acknowledge institutional support from Leibniz-IPHT and Friedrich Schiller University Jena. This work was supported in the innovation project ``Infrared single-photon detection via superconducting nanowires tailored by disorder'' by the Carl-Zeiss-Stiftung as part of the CZS Center QPhoton. The XRR facilities are partially supported by the European Regional Development Fund (ERDF) under the Project No. FKZ 2023 HSB 0025.
	\end{acknowledgments}
	
	\appendix
	
	\section{Ellipsometric modeling before oxygen plasma}
	\label{app:before}

	\begin{figure*}[t] 
		\centering 
		\nolinenumbers 
		
		\resizebox{0.95\textwidth}{!}{%
			\begin{minipage}{\textwidth}
				\centering
				
				\begin{minipage}{0.492\textwidth} 
					\centering 
					\includegraphics[width=\linewidth]{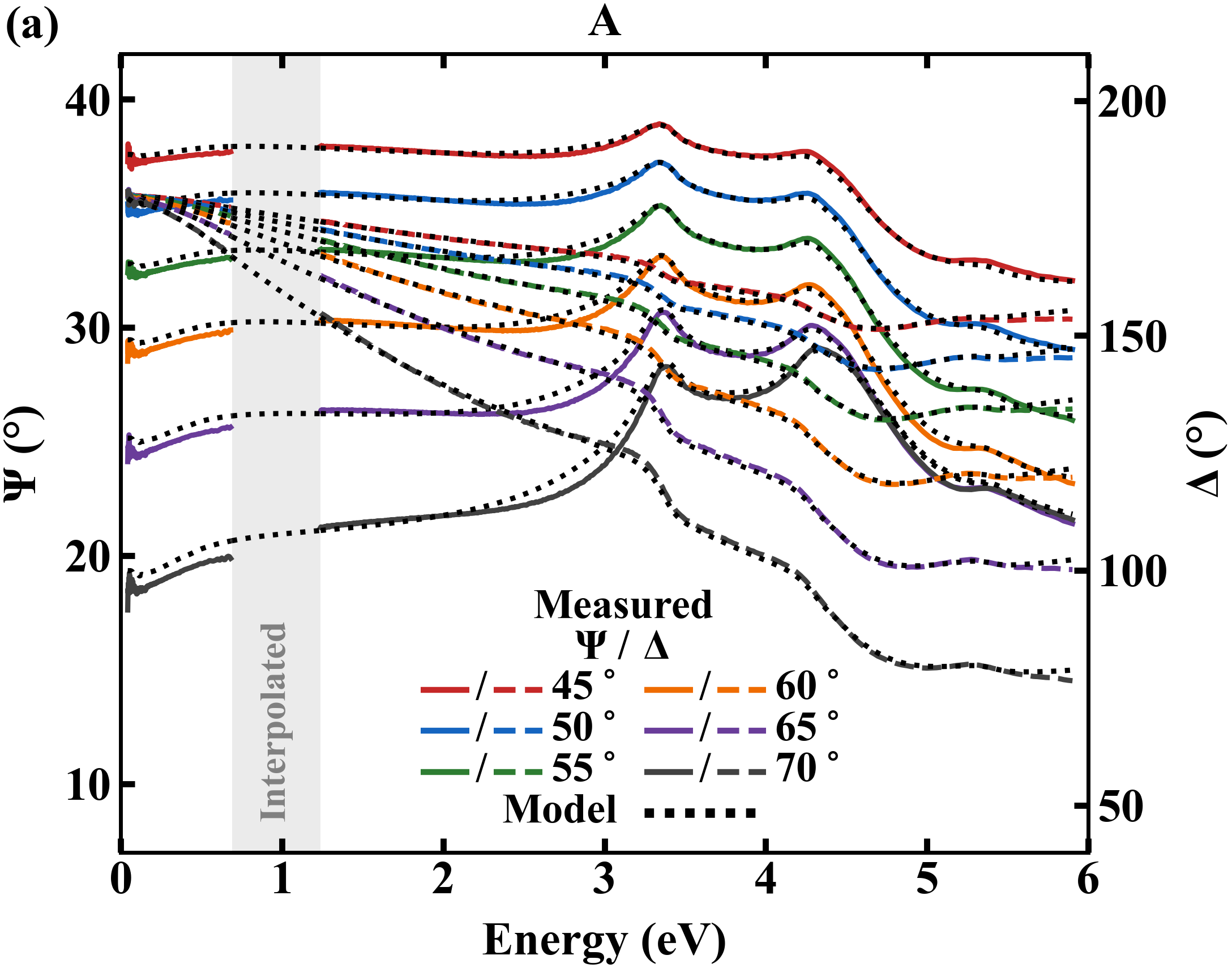} 
				\end{minipage}\hfill 
				\begin{minipage}{0.492\textwidth} 
					\centering 
					\includegraphics[width=\linewidth]{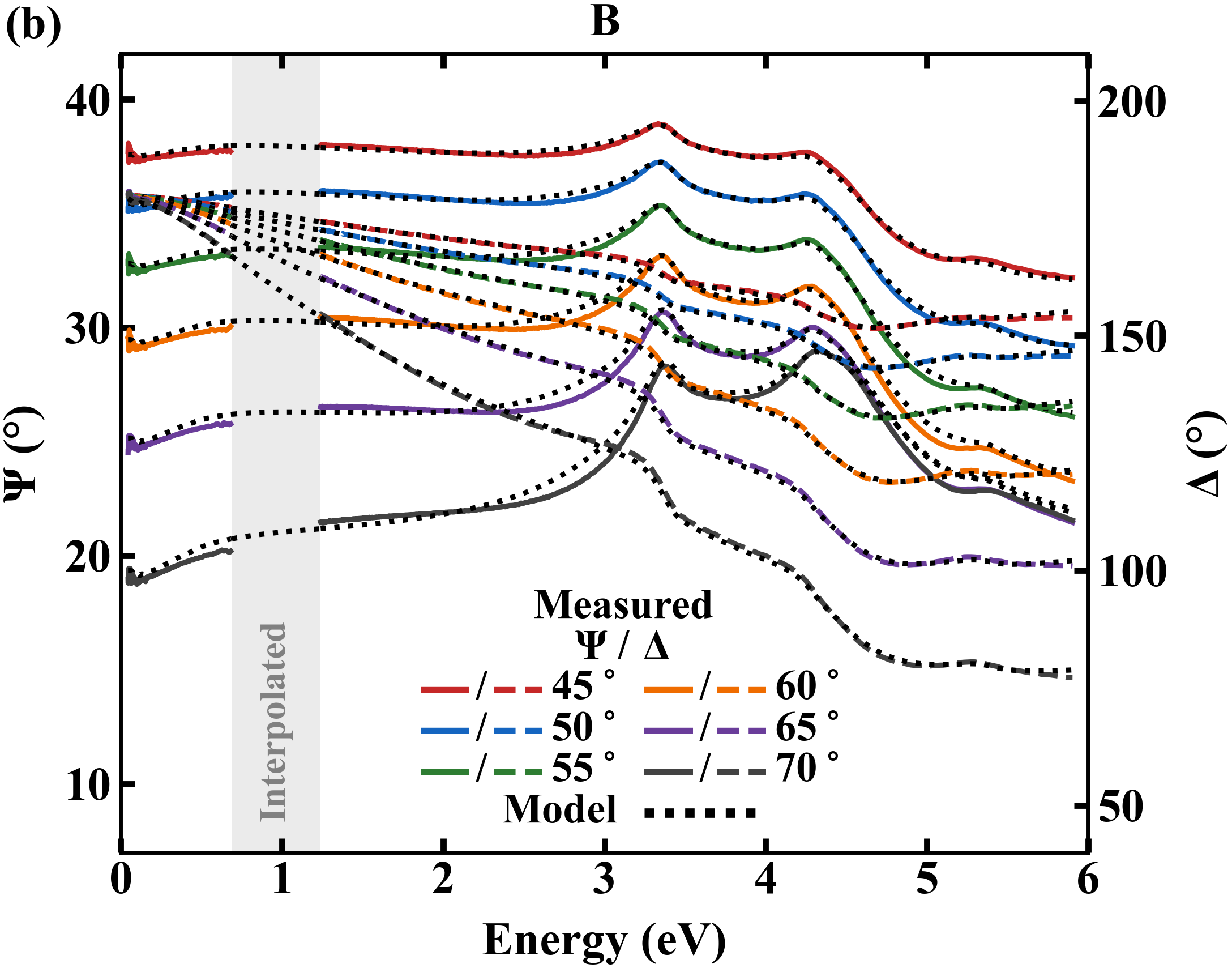} 
				\end{minipage} 
				
				\vspace{0.25cm} 
				
				\begin{minipage}{0.492\textwidth} 
					\centering 
					\includegraphics[width=\linewidth]{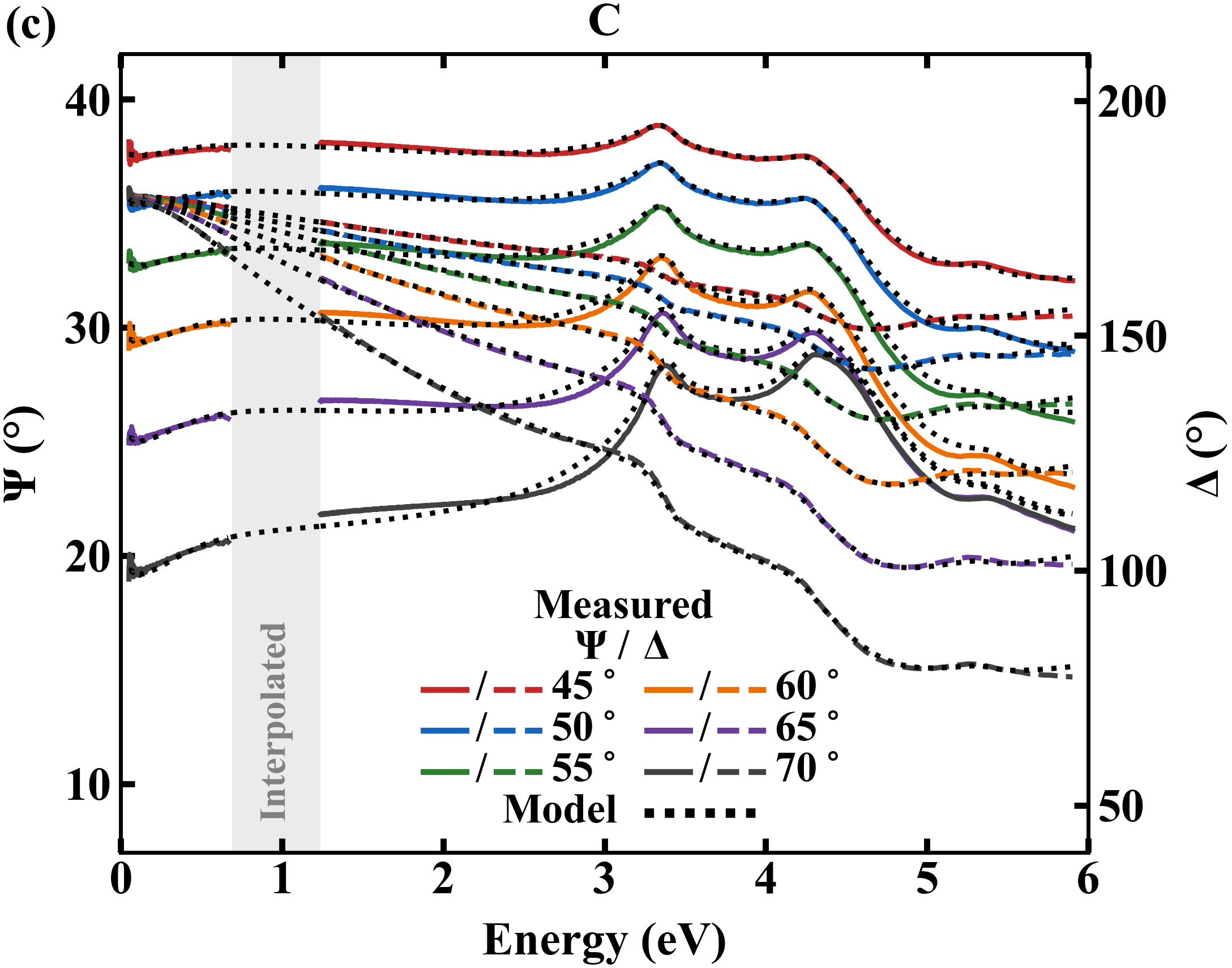} 
				\end{minipage}\hfill 
				\begin{minipage}{0.492\textwidth} 
					\centering 
					\includegraphics[width=\linewidth]{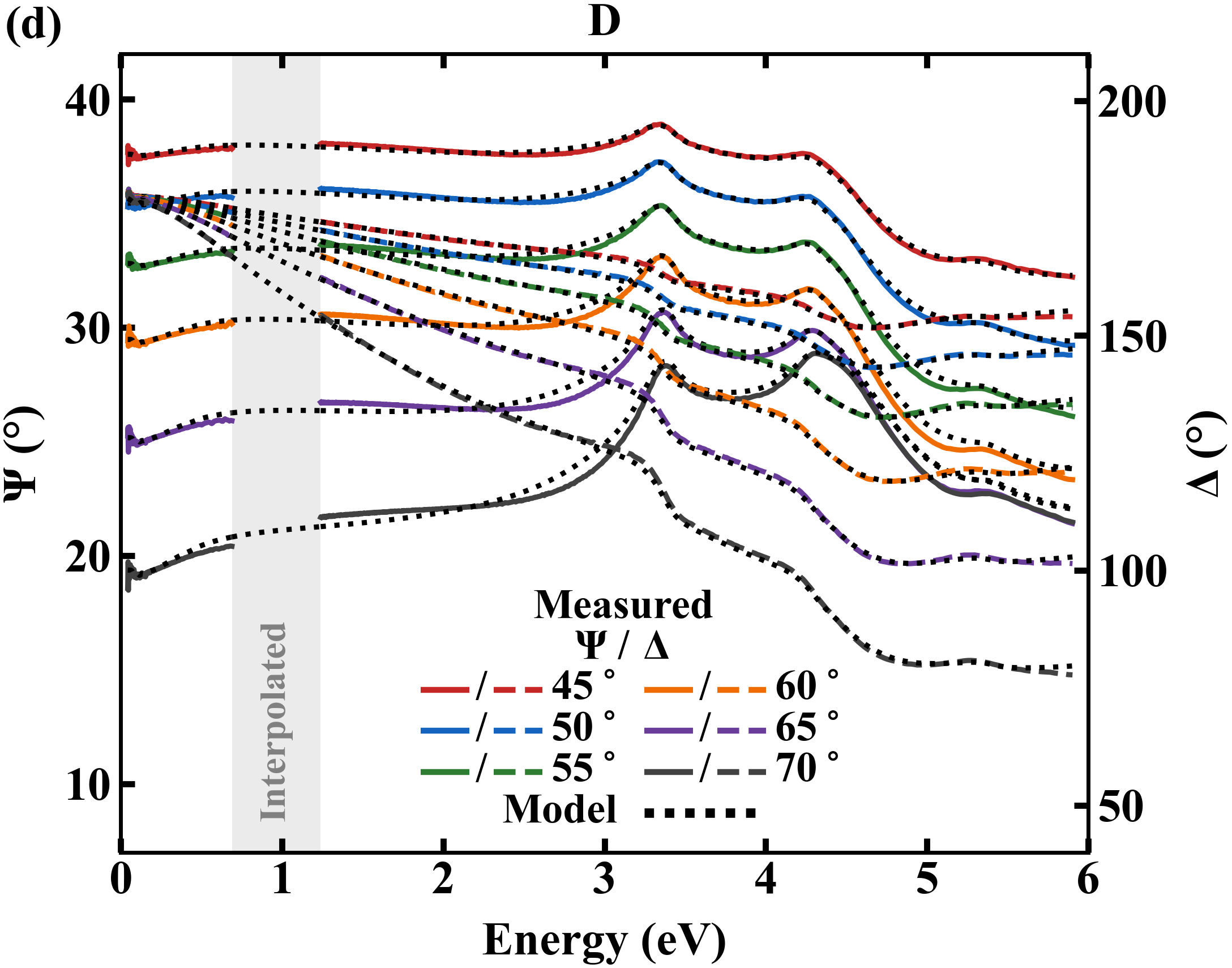} 
				\end{minipage}
				
			\end{minipage}%
		}
		
		\caption{Measured and modeled ellipsometric $\Psi$ and $\Delta$ spectra as a function of photon energy for the NbN thin films before oxygen plasma treatment: (a) A, (b) B, (c) C, and (d) D, acquired at incidence angles of $45^\circ$, $50^\circ$, $55^\circ$, $60^\circ$, $65^\circ$, and $70^\circ$. The spectral interval between 0.82 and 1.24\,eV (shaded in gray) was not measured directly because it lies between the spectral ranges of the UV--Vis and IR spectroscopic ellipsometers. The optical response in this interval was obtained by Kramers--Kronig-consistent interpolation within the dielectric-function model.} 
		\label{fig:psi_delta_before} 
	\end{figure*}
	
	\begin{table*}
		\caption{Best-fit Drude--Lorentz optical model parameters extracted from spectroscopic ellipsometry measurements of the NbN thin films prior to oxygen plasma treatment. Amp., Br., and En. refer to the oscillator amplitude, broadening, and resonance energy, respectively. The Drude parameters $N$, $\upmu$, and $m^*$ correspond to the charge-carrier density, mobility, and effective mass, while $\upvarepsilon_{\infty}$ is the high-frequency dielectric constant.}
		\label{tab:oscillator_parameters_before}
		\begin{ruledtabular}
			\begin{tabular*}{\textwidth}{@{\extracolsep{\fill}}lcccccccccccccc@{}}
				
				& \multicolumn{3}{c}{1\textsuperscript{st} UV Lorentz}
				& \multicolumn{3}{c}{2\textsuperscript{nd} UV Lorentz}
				& \multicolumn{3}{c}{IR Lorentz}
				& \multicolumn{3}{c}{Drude}
				& & \\
				
				Sample
				& Amp. & Br. & En.
				& Amp. & Br. & En.
				& Amp. & Br. & En.
				& $N$ & $\upmu$ & $m^*$
				& $\upvarepsilon_{\infty}$ & Roughness \\
				
				& & (eV) & (eV)
				& & (eV) & (eV)
				& & (eV) & (eV)
				& ($10^{22}$\,cm$^{-3}$)
				& (cm$^2$\,V$^{-1}$\,s$^{-1}$)
				&
				&
				& (nm) \\
				
				\hline
				
				A
				& 6.39 & 7.97 & 7.26
				& --- & --- & ---
				& 43.00 & 2.37 & 0.81
				& 1.21
				& 2.44
				& 1.82
				& 1.00
				& 0.10 \\
				
				B
				& 6.37 & 8.29 & 7.29
				& --- & --- & ---
				& 42.70 & 2.37 & 0.81
				& 1.20
				& 2.44
				& 1.82
				& 1.00
				& 0.29 \\
				
				C
				& 6.71 & 7.33 & 7.14
				& --- & --- & ---
				& 43.50 & 2.37 & 0.81
				& 1.21
				& 2.44
				& 1.84
				& 1.00
				& 0.50 \\
				
				D
				& 6.58 & 7.85 & 7.18
				& --- & --- & ---
				& 43.50 & 2.37 & 0.81
				& 1.21
				& 2.44
				& 1.84
				& 1.00
				& 0.50 \\
				
			\end{tabular*}
		\end{ruledtabular}
	\end{table*}

	\begin{figure}[t]
		\centering
		\nolinenumbers
		
		\includegraphics[width=\linewidth]{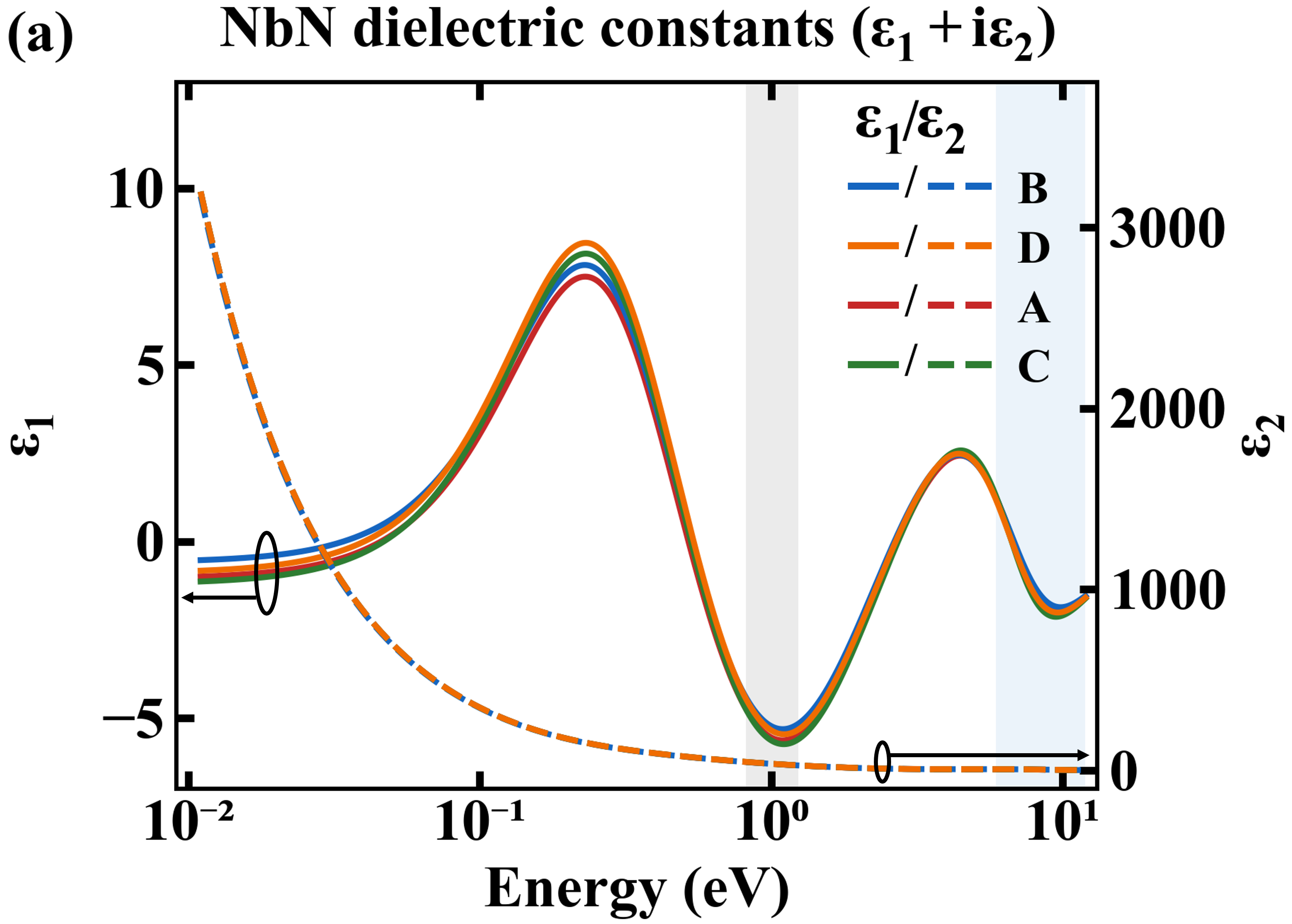}
		
		\vspace{0.25cm}
		
		\includegraphics[width=\linewidth]{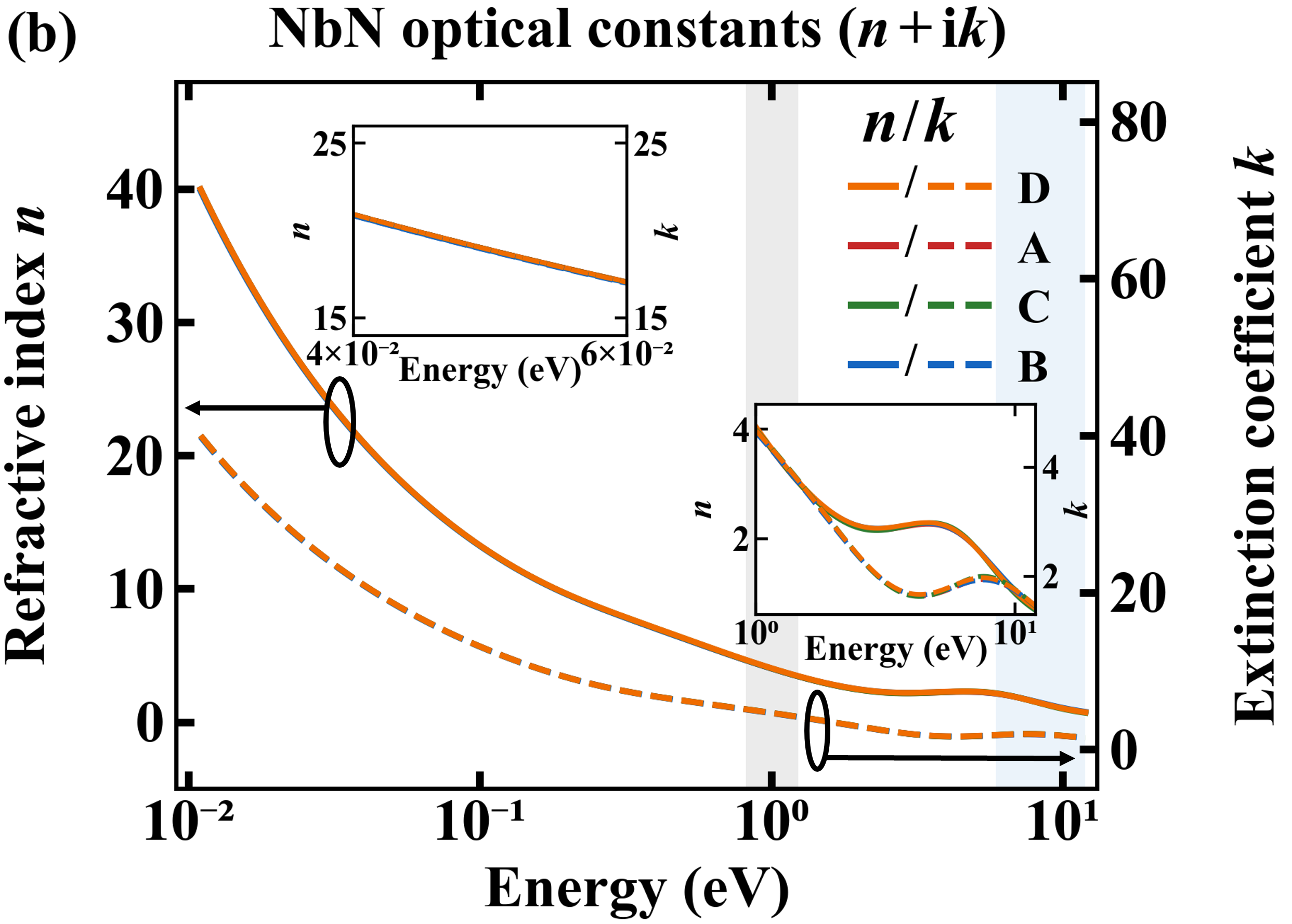}
		
		\caption{Dielectric function and optical constants of NbN thin films before oxygen plasma treatment. Panel (a) shows the $\upvarepsilon_1$ and $\upvarepsilon_2$ components of the dielectric function for samples A, B, C, and D, while (b) presents the corresponding refractive index ($n$) and extinction coefficient ($k$). The close similarity among the four spectra over most of the photon-energy range reflects the comparable optical response of the films before treatment. Enlarged views of the low- and high-energy regions in (b) allow a more detailed comparison of the small spectral differences that are less apparent in the full-range plots. The gray-shaded region from 0.82 to 1.24\,eV denotes the interpolated response, whereas the blue-shaded region from 6.20 to 12\,eV represents the extrapolated response, both maintaining Kramers--Kronig consistency within the same optical model.}
		\label{fig:dielectric_optical_before}
	\end{figure}
	
	The ellipsometric data obtained before oxygen plasma treatment were analyzed using the same multilayer structure as for the plasma-treated samples. The dielectric function of NbN was described by one UV Lorentz oscillator, one IR Lorentz oscillator, and a Drude term. An additional UV Lorentz oscillator was not required to describe the measured optical response. Figure~\ref{fig:psi_delta_before} compares the measured and calculated $\Psi$ and $\Delta$ values for samples A, B, C, and D at incidence angles from $45^\circ$ to $70^\circ$ in $5^\circ$ steps. Good agreement was obtained between the experimental data and the model calculations for all four samples. In the unmeasured spectral interval between 0.82 and 1.24\,eV, the dielectric response was calculated from the fitted Drude--Lorentz model, which satisfies the Kramers--Kronig relations. The fitted model parameters are summarized in Table~\ref{tab:oscillator_parameters_before}. The UV and IR Lorentz oscillator parameters, together with the Drude parameters, show only small variations among the four samples, indicating similar interband and free-carrier optical responses before oxygen plasma treatment. The high-frequency dielectric constant $\upvarepsilon_{\infty}$ was fixed at 1.00 for all samples, while the surface roughness was represented by an EMA layer and determined from the ellipsometric fit.

	The complex dielectric function derived from the fitted model is presented in Fig.~\hyperref[fig:dielectric_optical_before]{\ref*{fig:dielectric_optical_before}(a)}. The dielectric response of samples A, B, C, and D exhibits very similar spectral characteristics across the measured energy range, indicating that the electronic structure and optical properties of the untreated NbN films remained largely consistent. At low photon energies, the increase in $\upvarepsilon_2$ reflects the dominant free-carrier response characteristic of metallic NbN \cite{Kern2024, Sanjines2006}. In contrast, $\upvarepsilon_1$ approaches only slightly negative values toward the low-energy limit. For a good metal with weak carrier scattering, the Drude response would generally produce a much larger negative $\upvarepsilon_1$ in this regime \cite{Ordal1983}. The comparatively small negative values observed here are consistent with the relatively strong carrier scattering in the NbN films, which reduces the magnitude of the negative Drude contribution to $\upvarepsilon_1$. Thus, although the strong low-energy increase in $\upvarepsilon_2$ clearly reflects the metallic free-carrier response, the corresponding real part exhibits only a weakly negative Drude-like behavior. As the photon energy increases, the influence of the free carriers gradually decreases, while the dielectric function is governed primarily by interband electronic transitions. The optical constants, obtained from the complex dielectric function, likewise remain within a narrow range for the four samples (see Fig.~\hyperref[fig:dielectric_optical_before]{\ref*{fig:dielectric_optical_before}(b)}). The close overlap of the $n$ spectra indicates similar dispersive behavior, whereas the comparable $k$ spectra demonstrate that the optical attenuation and absorption characteristics of the films were closely matched before oxygen plasma treatment. Minor sample-to-sample differences become more apparent at higher photon energies, where the optical response is more sensitive to variations in the ultraviolet Lorentz parameters listed in Table~\ref{tab:oscillator_parameters_before}.
	
	\begin{figure*}[t] 
		\centering 
		\nolinenumbers 
		
		\resizebox{0.95\textwidth}{!}{%
			\begin{minipage}{\textwidth}
				\centering
				
				\begin{minipage}{0.492\textwidth} 
					\centering 
					\includegraphics[width=\linewidth]{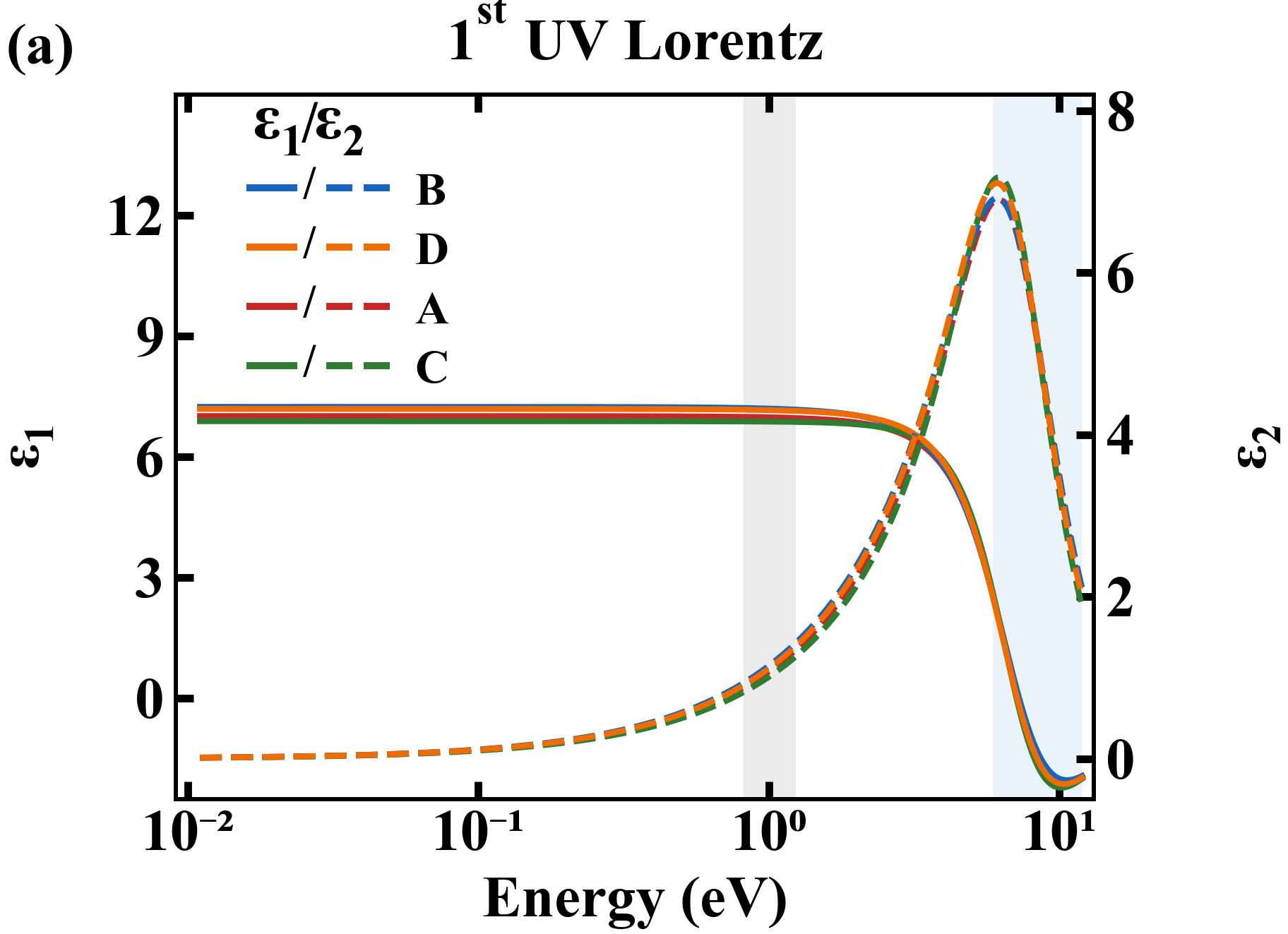} 
				\end{minipage}\hfill 
				\begin{minipage}{0.492\textwidth} 
					\centering 
					\includegraphics[width=\linewidth]{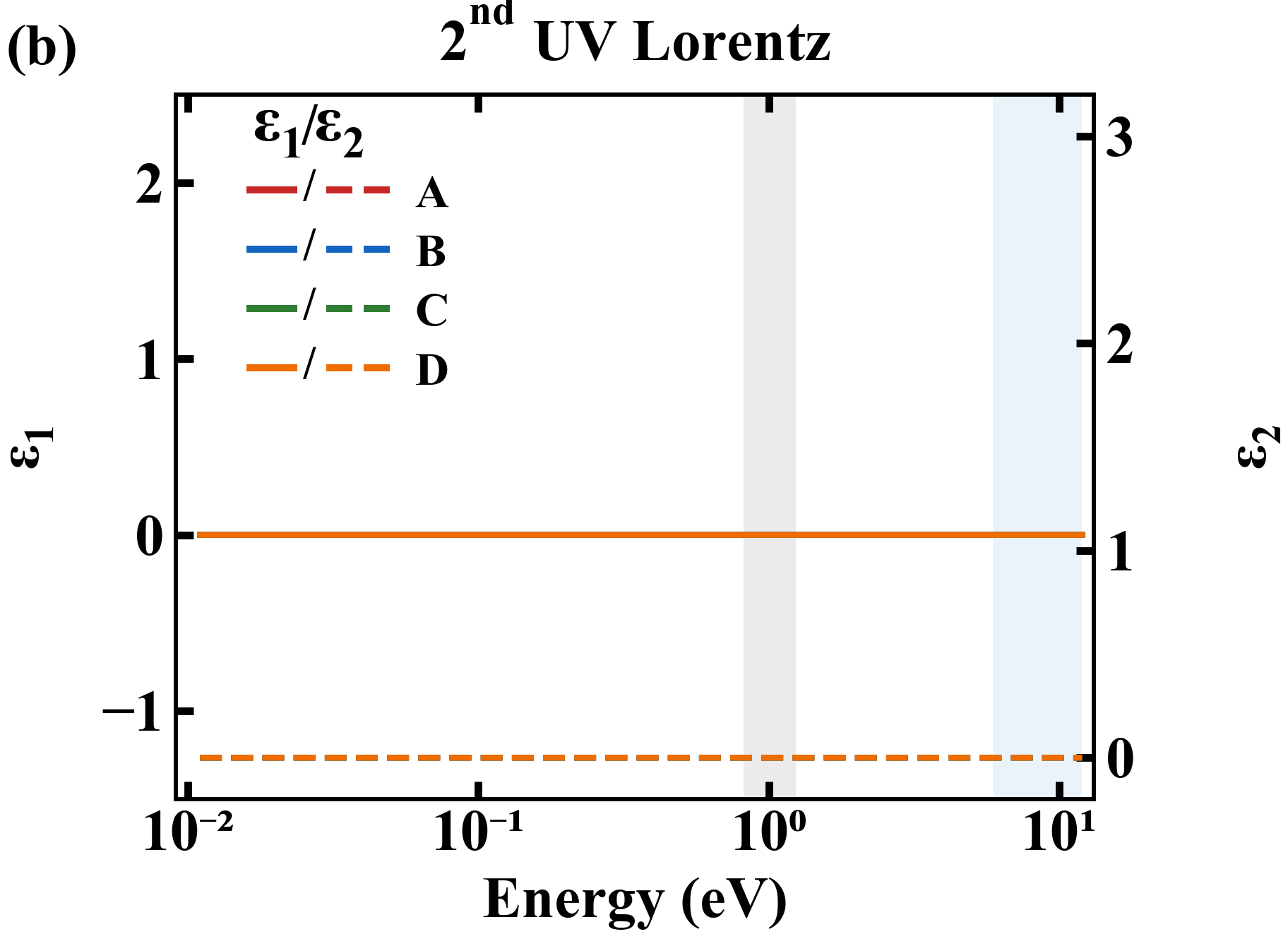} 
				\end{minipage} 
				
				\vspace{0.25cm} 
				
				\begin{minipage}{0.492\textwidth} 
					\centering 
					\includegraphics[width=\linewidth]{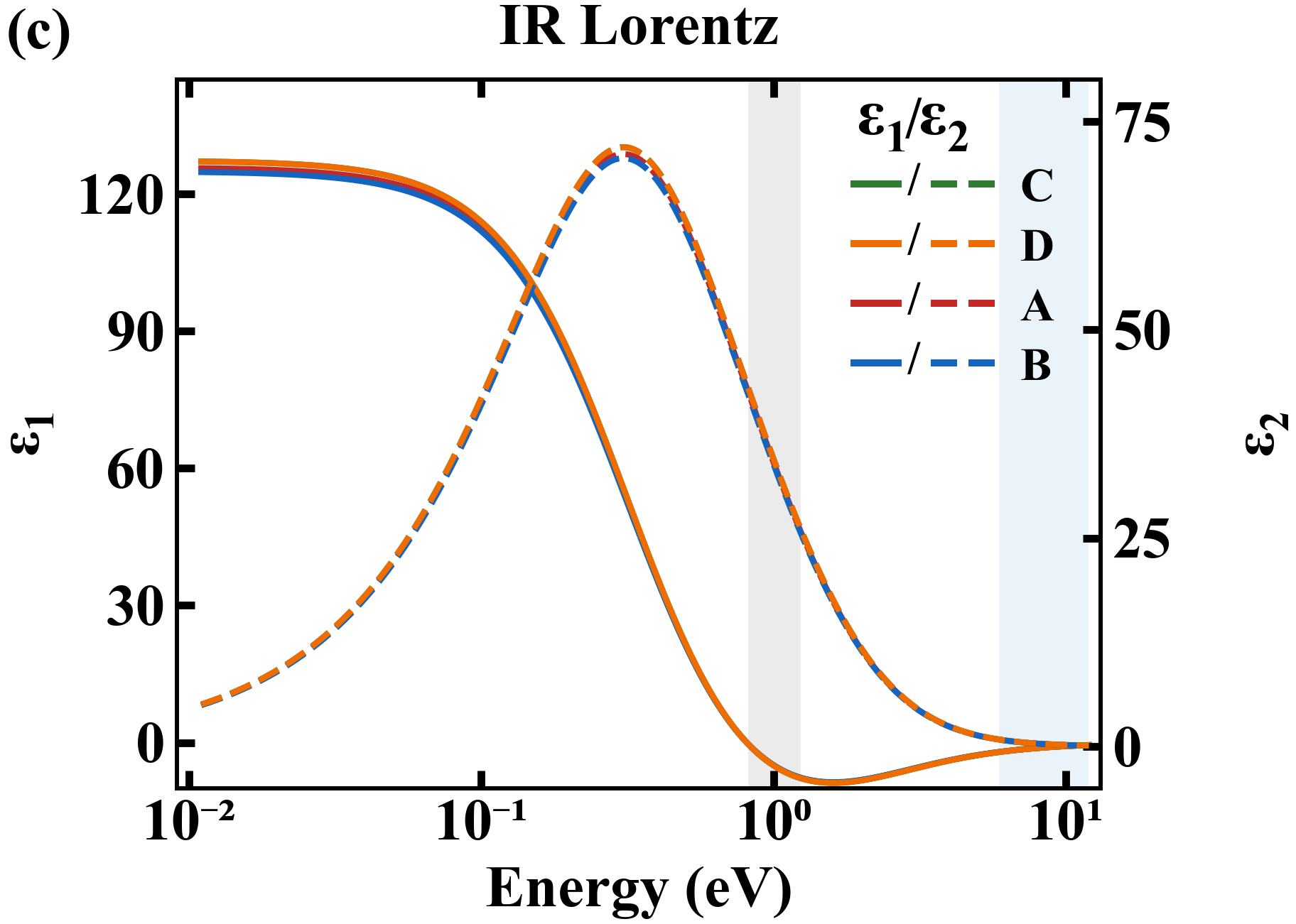} 
				\end{minipage}\hfill 
				\begin{minipage}{0.492\textwidth} 
					\centering 
					\includegraphics[width=\linewidth]{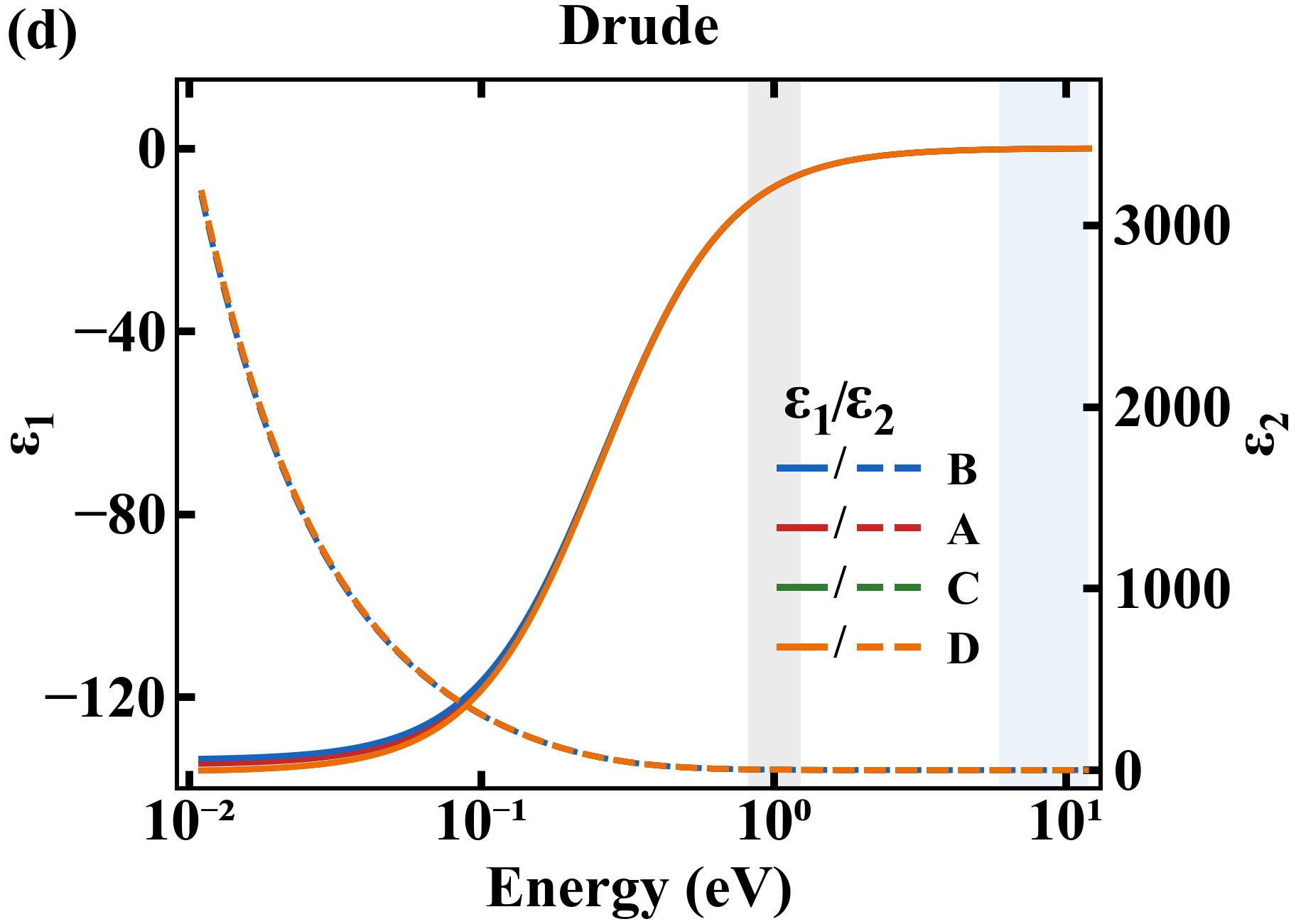} 
				\end{minipage} 
				
			\end{minipage}%
		}
		
		\caption{Dielectric-function contributions of the individual components in the Drude--Lorentz model for NbN thin films before oxygen plasma treatment: (a) the $1^{\mathrm{st}}$ UV Lorentz oscillator, describing the principal interband absorption of NbN associated with the strong hybridization of Nb 4$d$ and N 2$p$ electronic states; (b) the $2^{\mathrm{nd}}$ UV Lorentz oscillator, associated with oxygen-related electronic transitions, which was not required in the optical model before oxygen plasma treatment and therefore contributes neither to $\upvarepsilon_1$ nor $\upvarepsilon_2$; (c) the IR Lorentz oscillator, describing the low-energy anti-Drude-like optical response; and (d) the Drude term representing the free-carrier contribution. The gray-shaded region between 0.82 and 1.24\,eV denotes the interpolated response, whereas the blue-shaded region from 6.20 to 12\,eV represents the extrapolated response beyond the measured spectral range, both described by the same Kramers--Kronig-consistent optical model.} 
		\label{fig:oscillators_before} 
	\end{figure*}
	
	The individual contributions of the Drude--Lorentz model to the dielectric function are presented in Fig.~\ref{fig:oscillators_before}. The $1^{\mathrm{st}}$ UV Lorentz oscillator, shown in Fig.~\hyperref[fig:oscillators_before]{\ref*{fig:oscillators_before}(a)}, describes the high-energy interband response associated with hybridized Nb 4$d$ and N 2$p$ electronic states \cite{Kern2024,Farha2018,Sanjines2006}. The resonance energies range from 7.14 to 7.29\,eV, with small variations in oscillator amplitude and broadening among the four samples (Table~\ref{tab:oscillator_parameters_before}). In contrast, the $2^{\mathrm{nd}}$ UV Lorentz oscillator was not included in the model for the untreated samples [Fig.~\hyperref[fig:oscillators_before]{\ref*{fig:oscillators_before}(b)}]. This establishes a reference for distinguishing the additional absorption feature near 4.9\,eV observed after oxygen plasma treatment from the initial optical response of NbN. The near-infrared optical response of NbN cannot be fully captured by the Drude term because disorder gives rise to deviations from ideal free-carrier behavior \cite{Kern2024, Fratini2021, Kern2021}. This displaced Drude contribution is accounted for by the IR Lorentz oscillator in the dielectric-function model [Fig.~\hyperref[fig:oscillators_before]{\ref*{fig:oscillators_before}(c)}]. The resonance energy remains fixed at 0.81\,eV, whereas only small changes are observed in the oscillator amplitude. These results indicate that the displaced Drude contribution was largely preserved among the untreated NbN films. At low photon energies, the dielectric function is dominated by the free-carrier response described by the Drude term \cite{Schubert2004,Fomra2024}, illustrated in Fig.~\hyperref[fig:oscillators_before]{\ref*{fig:oscillators_before}(d)}. Both $\upvarepsilon_{1}$ and $\upvarepsilon_{2}$ exhibit only minor differences among the four samples, consistent with their nearly identical Drude parameters. As summarized in Table~\ref{tab:oscillator_parameters_before}, $N$ remains nearly constant at approximately $1.21 \times 10^{22}$\,cm$^{-3}$, while $\upmu$ is 2.44\,cm$^{2}$\,V$^{-1}$\,s$^{-1}$ for all four samples.

	\bibliography{references}
\end{document}